\documentclass[10pt,journal,IEEEtran]{IEEEtran}

\ifCLASSOPTIONcompsoc
  \usepackage[nocompress]{cite}
\else
  \usepackage{cite}
\fi
\usepackage[table]{xcolor}
\usepackage{amsthm}
\usepackage{mathtools}
\usepackage{multirow}
\usepackage{colortbl}
\usepackage{hyperref}
\usepackage{hypcap}
\hypersetup{colorlinks=true,allcolors=blue}

\ifCLASSINFOpdf

\else
 
\fi

\usepackage{color}
\usepackage{hhline}
\usepackage{csquotes}
\usepackage{enumitem}

 \usepackage{graphicx}
\begin{document}

\title{Communicating Using Spatial Mode Multiplexing: Potentials, Challenges and Perspectives}

\author{Abderrahmen Trichili, \emph{Member, IEEE}, Ki-Hong Park, \emph{Member, IEEE}, Mourad Zghal, Boon S. Ooi, \emph{Senior Member, IEEE}, Mohamed-Slim Alouini, \emph{Fellow, IEEE}
\IEEEcompsocitemizethanks{\IEEEcompsocthanksitem Abderrahmen Trichili, Ki-Hong Park, Boon S. Ooi and Mohamed-Slim Alouini are with the Computer, Electrical and Mathematical Sciences $\&$ Engineering in King Abdullah University of Science and Technology, Thuwal, Makkah Province, Kingdom of Saudi Arabia (email: \{abderrahmen.trichili,kihong.park,boon.ooi,slim.alouini\}@kaust.edu.sa).\newline

Mourad Zghal is with Gres'Com Laboratory, Engineering School of Communication of Tunis (SUP'COM), University of Carthage, Ariana, Tunisia (email: mourad.zghal@supcom.tn).

}
}

%


\IEEEtitleabstractindextext{%
\begin{abstract}
Time, polarization, and wavelength multiplexing schemes have been used to satisfy the growing need of transmission capacity. Using space as a new dimension for communication systems has been recently suggested as a versatile technique to address future bandwidth issues. We review the potentials of harnessing the space as an additional degree of freedom for communication applications including free space optics, optical fiber installation, underwater wireless optical links, on-chip interconnects, data center indoor connections, radio frequency and acoustic communications. We focus on the orbital angular momentum (OAM) modes and equally identify the challenges related to each of the applications of spatial modes and the particular OAM modes in communication. We further discuss the perspectives of this emerging technology. Finally, we provide the open research directions and we discuss the practical deployment of OAM communication links for different applications.
\end{abstract}

\begin{IEEEkeywords}
Space division multiplexing, orbital angular momentum, free space optics, optical fibers, vortex fibers, underwater wireless optical communication, radio frequency, millimeter waves, optical interconnects, optical switching, fiber nonlinearity, atmospheric turbulence, oceanic turbulence, crosstalk, adaptive optics.
\end{IEEEkeywords}}

\maketitle

\IEEEdisplaynontitleabstractindextext

%
\IEEEpeerreviewmaketitle



\section{Introduction}\label{sec:introduction}
The use of the spatial structure of the electromagnetic wave is mooted as a potential solution to address the pending $\emph{capacity crunch}$ \cite{RichardsonATPHO13}. In space division multiplexing (SDM) based communications, each spatial mode can act as an independent information-bearing carrier scaling the total transmission capacity by several orders of magnitude \cite{LiOPEX11}. Different mode shapes have been proposed to realize SDM in optical communication systems \cite{LiAOP14}. A particular mode set is orbital angular momentum (OAM) \cite{WillnerScience}. \newline
\indent OAMs were discovered by Les Allen and co-workers in 1992 \cite{AllenPRA92}. By definition an electromagnetic wave carrying an OAM of $\ell\hbar$ per photon is a wave possessing an azimuthally varying phase term $\exp(i\ell\phi)$, where $\ell$ is an unbounded integer known as the topological charge, $\phi$ is the azimuth and $\hbar$ is the Planck constant divided by $2\pi$. OAM beams posses a helical or twisted structure invisible to the naked eye but can be revealed through the interaction of light with matter. The OAM is sometimes ambiguous and should not be confused with the spin angular momentum (SAM) associated to circular polarization where the beam can only have two orthogonal states: right-handed or left-handed circularly polarized \cite{PadgetPRSA14}. This is different from twisted OAM beams, which can have an unlimited number of orthogonal states related to the charge or the commonly named azimuthal mode index $\ell$, which determines the number of wavefront helices for a light beam. $\ell$ may also be negative and its only difference besides the positive/negative value is the handedness. The use of this beam structure is found in various applications, ranging from imaging \cite{Furhapter,FooOL05}, quantum optics \cite{MairNAT01}, remote sensing \cite{RemoteSensing1,RemoteSensing2}, to optical trapping and manipulation\cite{ChenOL13, PadgettNATPHO11,PadgettOE17}. \newline
We know that several single or superpositions of orthogonal beams can carry OAM, either with a circular symmetry or an elliptic cylindrical symmetry, including Bessel beams \cite{BesselBeam}, Bessel-Gauss beams \cite{BesselGaussian}, Hermite Gaussian (HG) beams \cite{HermiteBessel}, Mathieu beams \cite{Mathieu}, Ince-Gaussian beams \cite{Ince}, vector vortex beams \cite{Vector}, and Laguerre Gaussian (LG) beams \cite{AllenPRA92}. Here, we mainly focus on the OAM modes derived from the LG mode family.\newline
\indent Since the seminal work of Gibson et al. in 2004 in \cite{GibsonOE04}, OAM has been implemented in free space optics (FSO), optical fibers, underwater wireless communication, indoor links, optical interconnects, and radio frequency (RF) communication. The OAM structure has been used not only for electromagnetic waves, but also for communication systems using acoustic waves.\newline
\indent There are two different ways of transferring information using OAMs. The first technique involves using OAM states, like letters, in a communication alphabet, which can be regarded as a direct modulation using the OAM indices. One mode can be transferred and decomposed at a time, or multiple modes can propagate together and get separated at the output. The second technique for transferring information is to encode an independent data stream on each carrier OAM mode to be multiplexed along with other modes, and transfer it over the communication channel. With the latter method, it is possible to combine OAM multiplexing, with wavelength division multiplexing (WDM) and polarization division multiplexing (PDM). Data can be additionally mapped onto high order modulation formats such as quadrature phase-shift keying (QPSK) and quadrature amplitude modulation (QAM).\newline
 Due to the orthogonality, the superposition of an OAM with another mode having the opposite topological charge allows the generation of a new mode with a petal-like transverse intensity profile orthogonal to other OAM states. The synthesis of OAM modes with distinct $\lvert\ell\lvert$ can further be used as independent data carrier orthogonal to the other states. \newline
An interesting way of using OAM states is multicasting communication. Data on a particular OAM state can be duplicated and sent to different users over many OAM channels, without decoding and re-encoding processes, and independently from the propagation medium or the application. 
\subsection{OAM Mode Generation and Detection}
\label{Sec:GenDet} 
Extensive work has been done on the generation and detection of the OAM modes. Note that both generation and detection depend on the nature of the wave, i. e. the lightwave, the RF wave or the acoustic wave. Here we expose the generation technique of lightwave OAMs in free space.\newline
OAM modes can be generated using different techniques including spiral phase plates (SPPs) \cite{OAMGeneration1,OAMGeneration2}, computer generated holograms (CGHs) programmed on spatial light modulators (SLMs) \cite{OAMGenerationSLM1,OAMGenerationSLM2}, integrated compact devices based on silicon resonators\cite{OAMGeneration4}, metamaterials \cite{OAMGeneration5,OAMGeneration6}, metasurfaces \cite{OAMGeneration7,OAMGeneration8}, and q plates \cite{Qplate1,Qplate2}. Spiral phase plates and q-plates are usually static optics; each SPP or q-plate allows the generation of a single OAM mode. SLMs are controllable devices that can generate a wide range of OAM modes by modulating the phases of Gaussian beams. It is also possible to alter the amplitude of Gaussian beams using phase-only SLMs \cite{Arrizon}. However, the refreshment image rate of most of commercially available SLMs is limited to 60 Hz which can be slow for some communication applications. A new technique for the fast generation of OAM has been proposed, using digital micro-mirror devices (DMDs) \cite{OAMGenerationDMD1,OAMGenerationDMD2,OAMGenerationDMD3}.\newline
\begin{figure*}[!t]
\centering
\includegraphics[width=5.2in]{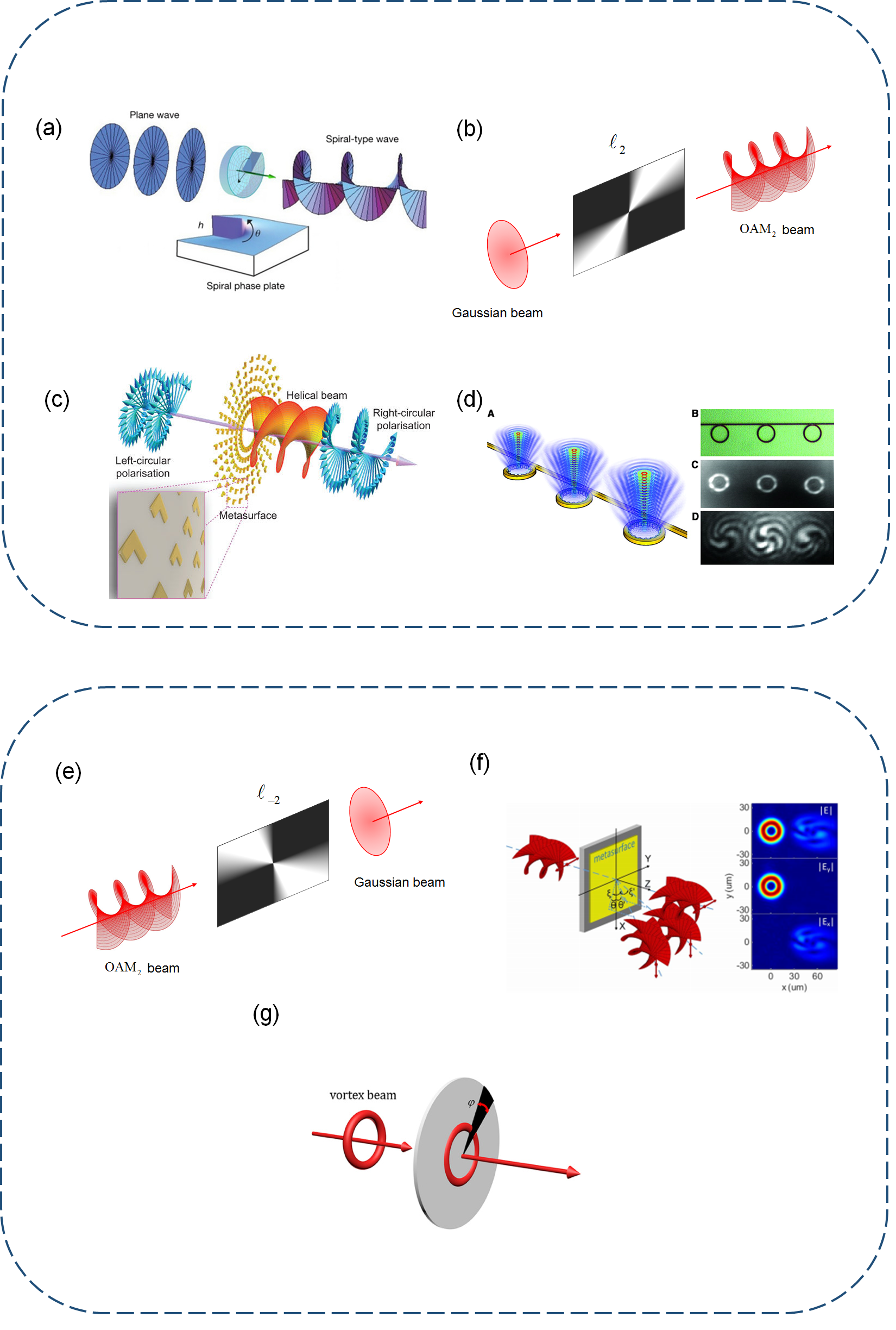}
\caption{Schematic illustration of different generation of OAMs and detection techniques. (a) Generation of OAM using a spiral phase plate \cite{OAMGeneration2}; (b) Generation of an OAM using computer-generated hologram; (c) Schematic representation of spin-to-orbit coupling through a metasurface \cite{OAMGeneration8}; (d) Illustration of an array of OAM emitters; (e) Detection of OAM using a digital hologram; (f) Measurement of the topological charge of OAMs using a metasurface \cite{OAMDetectionMeta1}; (g) Detection of an OAM mode using a sectorial screen \cite{OAMDetectionScreen}. }
\label{OAMGenDet}
\end{figure*}
\indent OAM light beam detection can be performed using several techniques including those used for the generation. By dynamically changing the hologram displayed on the SLM screen, it is possible to detect the topological charge of an unknown OAM beam and recover it back to a Gaussian shape \cite{Mdecomposition1,Mdecomposition2}. The idea is to perform an inner product measurement between an incoming OAM field with a match filter of $\exp(i\ell\phi)$, for various $\ell$ values. Plasmonic metamaterials have been proposed to efficiently detect OAM light beams in a broad spectrum, and to offer the capability of integrating them with fibers, on-chip devices as reported in \cite{OAMDetectionMeta1}. Methods including transformation optics \cite{BerkhoutPRL10,MirhosseiniNATCOM13} have also been shown to be valuable alternatives. In addition, a mode detection approach based on a  photonic silicon-integrated circuit is proposed in \cite{OAMDetectionPCircuit}. Authors of \cite{3DOAMDetector} fabricated a 3D micro-scale broadband mode sorter that is able to detect single OAM beam and de-multiplex superpositions of OAM beams with topological charge in the range of $-3\le\ell\le3$. Another detection method, based on a sectorial diffraction screen, was also reported in \cite{OAMDetectionScreen}, as low-cost option that is relatively easy to implement. Recently, machine-learning-based approaches have been implemented in order to accurately identify OAM modes, after their propagation in free space \cite{OAMDetectionMLearning,OAMDetectionPRecognition}; they offer great potential in mode detection without need for atmospheric turbulence mitigation (see details in section \ref{MachineLearning}). \newline
Various OAM light beams generation and detection techniques are presented in Fig.~\ref{OAMGenDet}. We note that most of the reviewed generation and detection techniques demonstrated in free space are also suitable for optical fibers.\newline
\indent In RF, mode converters, such as spiral phase plates, have been used to generate radio OAM beams \cite{OAMRadioSPP1,OAMRadioSPP2}. Another similar method uses a flat phase plate with different permittivities along its surface \cite{Flatplate}. Twisted parabolic reflectors have been used to generate OAM beams, as well \cite{ParabolicOAMReflector1,ParabolicOAMReflector2,ParabolicOAMReflector3}. Gui and his co-workers fabricated two circular slot antenna systems to generate OAM beams at 2.4 GHz \cite{SlotOAM}. Spiral antennas have been also used for the generation of circularly polarized OAM modes, as reported in \cite{OAMRFSASAA}. Metasurfaces have been also adopted to generate OAMs and equally to measure the topological charge of mixed $\ell$ values beams \cite{OAMRadioMeta,OAMDetectionMetaRF}. Circular and ring-shape arranged arrays of patch antennas were also used to generate and detect various OAMs with different charges \cite{OAMPatch1,OAMPatch2,SpinelloARXIV,RadialUCA}. Zhao et al. performed simulation studies to evaluate the performance of different arrangements of antenna arrays and found that antennas with circular arrays performed better than ring antenna arrays in terms of generated mode purity and power loss \cite{ZhaoGLOBECOM17}.\newline
\indent Acoustic OAM waves were generated using an array of speakers forming a multi-ring pattern and controlled using a digital processor while the detection is performed using an array of sensors in a 4 concentric rings arrangement.

\subsection{OAM Deployment}
\label{deployement}
Different OAM communication scenarios have been successfully demonstrated. However, before moving from laboratory testbed and proof-of-concept (PoC) experiments, several issues need to be addressed. The deployment of OAMs faces challenges depending on several factors strongly related to the propagation media. For example, free space OAM communication can be hindered by absorption, scattering and atmospheric turbulence-induced fading. Preserving the twist of OAMs, when propagating over fibers, is a major challenge when installing OAM optical fibers. Underwater conditions are the greatest concern for submerged laser communication. Mitigating turbulence effects for optical communication is key to developing OAM channels with high-bit rates. On-chip scale OAM requires proper mode generation and detection technology. For RF systems, designing proper antennas that overcome propagation effects, as well as divergence, can facilitate the implementation of radio OAM. In this article, we focus on the main concerns of spatial mode multiplexing, with an emphasis focusing on OAM modes; we identify a possible solution that has the potential to bring this technology from experimental test benches to an industrial setting.
\subsection{Related Reviews}
The potential of using multiple modes in communication, including OAMs, is discussed by several communication specialists in \cite{LiAOP14,WillnerAOP15,WillnerIEEECM,FSOReview,WillnerPTRSA}. In \cite{LiAOP14}, Li et al. focus on space division multiplexing in fibers, and suggest that the OAM mode set is a promising potential information carrier in future communication networks. Progress in OAM generation/detection techniques as well as in the applications of OAMs in free space and optical fibers are reviewed in \cite{WillnerAOP15}. Additionally, a short article by Willner provides a general overview of the potentials of twisted light in communication \cite{WillnerIEEECM}. The use of OAM modes, as an addition to Gaussian mode-based free space channels, is discussed in \cite{FSOReview}.\newline
\indent Here, we provide a comprehensive review of the potentials of OAM in various communication applications based on the most significant results reported in the literature. We analyze relevant challenges associated with mode multiplexing, and we discuss the prospect of using OAM for data transmission, in the future. We also summarize the open problems and the future research directions. A discussion of the practicability and cost of multi-OAM communication is also included. 


\section{OAM Potentials}
\subsection{Free Space Optical Communication}
\begin{figure}[!t]
\centering
\includegraphics[width=3in]{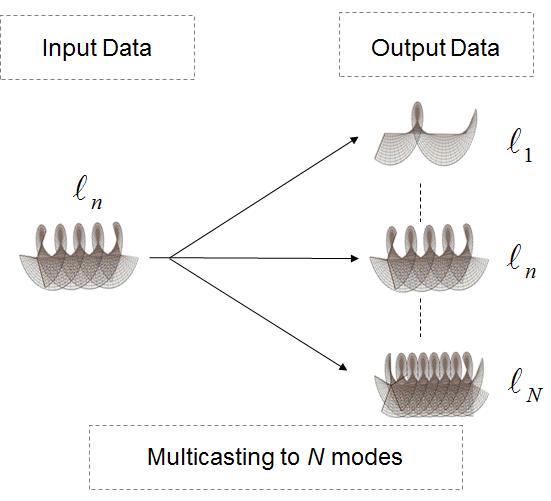}
\caption{Concept of the multicasting function in an OAM multiplexing system from one mode to $N$ modes.}
\label{MulticastingOAM}
\end{figure}

FSO is a license free wireless communication configuration that has recently received much interest for a variety of applications. FSO is an attractive solution for last mile connectivity problems, particularly for communication networks, when the installation of fiber optics is costly or not possible \cite{FSOApps}. FSO can be also used to establish inter-building secure communication, and can be deployed as a backup to optical fibers. Wireless optical communication can guarantee a line-of-sight (LoS) high bit rate wireless transmission over long distances, up to several kilometers. Furthermore, FSO communication is considered as a promising technique to scale down bandwidth challenges in future 5G networks \cite{5G}.\newline
\indent Multiple wavelength FSO provides better transmission, as demonstrated in \cite{FSO1,FSO2}. Data can be mapped on advanced modulation formats to achieve high bit rates and high spectral efficiency levels\cite{FSO3,FSO4,FSO5}.\newline
Another option is multiple-input and multiple-output (MIMO) FSO communication in which multiple lasers are positioned to transmit Gaussian beams to multiple receiving apertures \cite{MIMOFSO}. Over the last few years, it has been proven that it is possible to transmit information over spatially structured light beams \cite{GibsonOE04,Lavery280G,VectorFSO} including OAM beams and plane waves. Our focus is on the OAM modes of light.\newline
\indent The first PoC communication experiment incorporating OAMs in free space was carried out in 2004 by Gibson and co-workers \cite{GibsonOE04}. Over a 15-meter long link, 8 $\ell$-spaced values were chosen along the Gaussian beam as an alphabet for the communication.
OAM beams were generated and detected using two light modulators.
Since then, much progress has been made and free space transmission capacity of over 1 Tbit/s is, today, possible \cite{WangNATPHO12,2TbitOPEX,Massive}. By performing three dimensional multiplexing, Huang and co-workers were able to attain a 100 Tbit/s free-space data transmission using 12 OAM channels over 2 orthogonal polarization states and 42 wavelengths \cite{OAMFSO}. Very high spectral-efficient communication was reported by Wang et al. \cite{WangOFC}. More recent studies (2014) by the same authors reported free-space communication with a bit rate of 1.036 Pbit/s and a spectral efficiency of 112.6 bit/s/Hz, using 26 OAM states \cite{WangECOC14}. A year later, an ultra-high spectral efficiency of 435 bit/s/Hz was also reported by Wang et al., who used a $N$-dimensional multiplexing and modulation link with dual-polarization 52 OAM modes carrying Nyquist 32-QAM signals \cite{WangECOC15}. Such levels of transmission capacity were never attained before, in a laboratory setting, with any communication technique. These results only re-emphasize the high level of expectation from multimode free-space links.\newline
\indent The use of OAM modes as a communication alphabet has been also performed. In 2014, Krenn et al. reported a 3 km communication using multiple superpositions of OAM modes through turbulent air across Vienna \cite{KrennNJP14}. Although the main objective of their study was not to realize a high bit rate communication, the authors successfully demonstrated the ability of generating and detecting multiple OAMs after propagating them over a long distance, in a turbulent weather. In 2016, Br{\"u}ning et al. established a communication link using multiple azimuthal modes through a few mode fiber and a free space channel \cite{ForbesJO16}. Their studies were also performed using SLMs. We recall that SLMs allow the generation and detection of single and multiple OAM states, using computer-generated holograms \cite{DuOL15,ForbesAOP16}.\newline
\indent Additional experimental studies revealed the possibility of carrying encoded information, with a four-level pulse-amplitude modulation (PAM-4), using free space OAM modes \cite{LiuOPEX16}. OAM was also applied in a data multicasting transmission, as reported in \cite{MulticastingFSO}, where a 100 Gbit/s QPSK data signal carried by one OAM was duplicated onto 5 and 7 equal-spaced OAM transmissions. The concept of OAM multicasting is illustrated in Fig.~\ref{MulticastingOAM}.\newline
\indent Sun and Djorjevic numerically investigated the physical layer security (PLS) of OAM-based FSO links subject to atmospheric turbulence. The authors found that under weak and moderate turbulence regimes OAM-based FSO systems could provide higher security communications compared to single-mode FSO systems \cite{OAMPLS}.
Single mode-based products for FSO communication, are commercially available, offering transmission capacities up to few Gbit/s. In Table \ref{FSOProducts}, we present the specifications of the available FSO devices. If such devices are combined with OAM modes, higher communication throughputs can be achieved, paving the way for applications requiring high data rates such as 3D video conferencing. 
\begin{table}[!t]
\renewcommand{\arraystretch}{1.3}
\caption{Specifications of commercially available FSO systems}
\label{FSOProducts}
\centering
\begin{tabular}{|>{\columncolor[gray]{0.8}}l|l|l|l|}
\hline
\rowcolor[gray]{0.8}
Company&Model &Throughput & Range\\
\hline
Northen Storm \cite{NSProduct}&NS10G&10 Gbit/s&50-1500 m\\
\hline
Mostcom \cite{MostcomProduct}&Artolink&10 Gbit/s&2.5 km\\
\hline
CBL \cite{CBLProduct}&IP1000 plus&1 Gbit/s&800 m\\
\hline
JV Labs \cite{JVProduct}&iRedStar 1Gbps&1 Gbit/s&400 m\\
\hline
\end{tabular}
\end{table} 
\subsection{Optical Fiber Communication}    
\label{FiberPotential}                                                                
\begin{figure}[!t]
\centering
\includegraphics[width=3in]{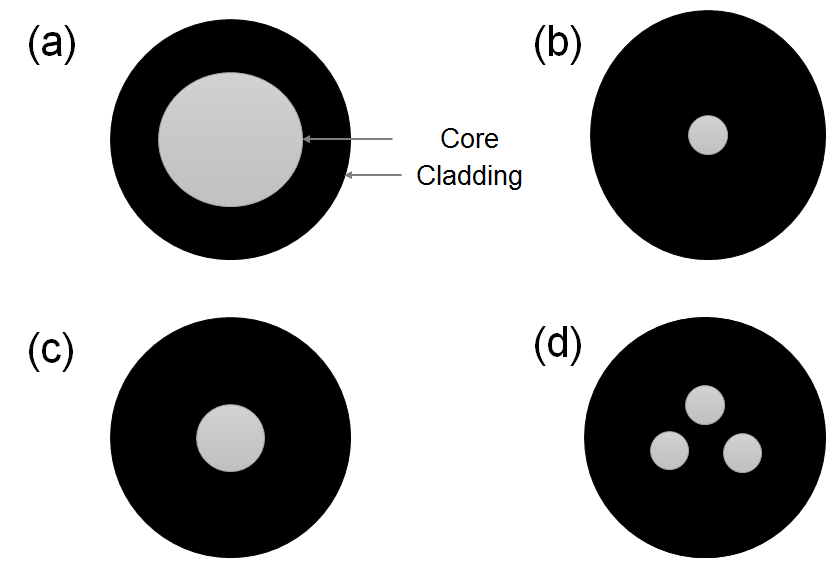}
\caption{A schematic illustrating different kind of optical fibers; (a) Multimode fiber, (b) single mode fiber, (c) few mode fiber and (d) multicore fiber}
\label{Fibers}
\end{figure}

Multimode fibers (MMFs) were extensively studied in the 1980s \cite{Berdague,Marcuse} with a primary focus on the coupling mechanisms between modes and the inter-modal energy distribution during propagation. The idea of using MMF was later abandoned, due to severe inter-modal dispersion limitations. Conventional MMFs have large cores that are usually approximately 50 $\mathrm{\mu m}$ and can support hundreds of modes were replaced by single mode fibers (SMFs) that have a relatively small core radius not exceeding 10 $\mathrm{\mu m}$. The transverse profiles of an MMF and an SMF are depicted in Fig.~\ref{Fibers}(a) and Fig.~\ref{Fibers}(b). However, MMFs have been recently reintroduced, for a few mode fibers (FMFs) having a relatively small core radius (See Fig.~\ref{Fibers}(c)); they support a limited number of modes, as one of the key components for SDM for optical networks \cite{RichardsonSC10}. The general idea of SDM, when first proposed, is to use optical fibers that support few spatial modes or multiple cores \cite{RichardsonATPHO13}. The latter fiber type is called a `multicore fiber' (MCF) and can be simply viewed as a superposition of multiple SMFs sharing the same fiber cladding (See Fig.~\ref{Fibers}(d)). The maximum data transfer capacity of MCFs directly scales with the number of cores used, and therefore should be carefully spaced inside the cladding to avoid crosstalk. A transmission capacity exceeding 100 Tbit/s over a 7-core fiber has been reported \cite{MulticoreFiber}. The potential of MCFs in optical communication is reviewed in details in \cite{Survey}.\newline
\indent Launching multiple modes over the same FMF is better known as `mode division multiplexing' (MDM) which was firstly adopted by Berdagu\'e and Facq to describe launching and retrieving two modes in a 10-m long MMF through spatial filtering techniques in 1982 \cite{Berdague}. The naming MDM is also equivalent to spatial mode multiplexing since it refers to launching multiple modes, over the same wavelength and on the same polarization state, simultaneously into the same medium. The system capacity of an MDM sytem is directly related to the number of simultaneously co-propagating spatial modes.
One of the first pioneering papers, to experimentally demonstrate MDM-based communication, reported a 33-km transmission of 28 Gbaud-QPSK signals over 6 spatial modes \cite{RandelFMF}. A $6\times6$ MIMO digital signal processing (DSP) was used to compensate for the inter-modal energy exchange between co-propagating modes. \newline
\indent Combining the two SDM approaches is also possible and has led to high transmission throughputs up to 200 Tbit/s over a 1-km long few mode multicore fiber \cite{FMFMCF}.\newline
\indent The most commonly used modes for fibers are known as `linearly polarized' (LP) modes. LP modes are not exact fiber modes, and can be simply viewed as combinations of fiber eigenmodes transverse electric (TE), transverse magnetic (TM), hybrid electric modes of EH and HE types. An LP mode can only propagate under the weakly guiding approximation which assumes that the difference between the fiber core and cladding refractive indices is very low. Since they are not exact solutions for fibers, linear coupling may occur between mode groups during propagation. MIMO DSP is needed to retrieve the data initially encoded on each mode, possibly resulting in demultiplexing complexity, especially for short length fiber links where MIMO modules are not required to compensate for chromatic dispersion \cite{RuschIEEECM}. \newline 
\indent For MDM, avoiding inter-modal coupling interactions and decreasing MIMO demultiplexing complexity requires the use of specific optical modes. OAM was proposed as a good option \cite{RuschIEEECM}. OAM fiber modes can be denoted as OAM$_{\ell,m}$, where $\ell$ is the azimuthal index and $m$ is the number of the concentric radial rings in the transverse intensity profile of the modes. OAM fiber modes can be expressed as functions of hybrid HE or EH modes \cite{OAMFiberModes}:
\begin{equation}
\begin{split}
&\mathrm{OAM}_{\pm\ell,m}^{\pm}=\mathrm{HE}_{\ell+1,m}^{\mathrm{even}}\pm i\mathrm{HE}_{\ell+1,m}^{\mathrm{odd}}\\
&\mathrm{OAM}_{\pm\ell,m}^{\mp}=\mathrm{EH}_{\ell-1,m}^{\mathrm{even}}\pm i\mathrm{EH}_{\ell-1,m}^{\mathrm{odd}},
\end{split}
\end{equation}
where subscripts refers to the handedness of the rotation of the circular polarization states.\newline
$\mathrm{HE}_{\ell+1,m}^{\mathrm{even}}$ ($\mathrm{EH}_{\ell-1,m}^{\mathrm{even}}$) and $\mathrm{HE}_{\ell+1,m}^{\mathrm{odd}}$ ($\mathrm{EH}_{\ell-1,m}^{\mathrm{odd}}$) are degenerate (have the same effective refractive index) which means that the linear combination modes $\mathrm{OAM}_{\pm\ell,m}^{\pm}$ ($\mathrm{OAM}_{\pm\ell,m}^{\mp}$) are also eigenmodes of the fiber and can be used as independent data carriers.\newline
\indent The very first paper to investigate the propagation of OAM modes in optical fibers was published in 2009 by Ramachandran et al.. In this paper, the authors reported the generation of optical vortices in fibers \cite{RamachandranOL09}. Three years later (in 2012), the same team highlighted the potential of a new kind of fibers, known as vortex fiber, that  preserves the twist of OAM modes while propagating \cite{RamachandranECOC12}. The same 1.1 km-long fiber was used to achieve the first terabit-scale date transmission, as reported in \cite{Bozinovic}. In this study, an aggregated transmission capacity of 1.6 Tbit/s was reached using two OAM modes over 10 wavelength channels around 1550 nm. Because OAM modes are orthogonal, MIMO DSP, in this case, is not crucial to get an efficient demultiplexing of OAM, contrary to LP mode multiplexing \cite{RuschIEEECM}. A full-duplex transmission capability using OAM modes over a vortex fiber was reported in \cite{ChenECOC16}.\newline
\indent In \cite{HuangSREP15}, Huang et al. reported the successful transmission of OAM modes over a 5 km long-FMF.
The possibility of transmitting OAM modes over large core conventional graded index MMFs was analyzed theoretically in \cite{OAMFiberAnaly}, and experimentally reported in \cite{ConventionalMMF1,ConventionalMMF2,ConventionalMMF3,OAMFMFMMF}. Ingerslev et al. reported the successful transmission of 12 modes which is the highest reported number of modes, so far transmitted in a MIMO-free configuration beyond 1 km of propagation \cite{12ModeOPEX}.
Table \ref{OAMFIBER}, summarizes the reported experimental demonstrations of OAM fiber transmissions, and highlights the key contributions brought by each study.\newline
\begin{table*}[!t] 
\renewcommand{\arraystretch}{1.3}
\caption{Summary of the literature results on the OAM communications over optical fibers}
\label{OAMFIBER}
\centering
\begin{tabular}{|>{\columncolor[gray]{0.8}}l|l|l|l|p{0.45\textwidth}|}
\hline
\rowcolor[gray]{0.8}
Reference&Year &Propagation Distance &Throughput&Main Contributions\\
\hline
\cite{Bozinovic}&2013&1.1 km&1.6 Tbit/s&\vspace{-\topsep}\begin{itemize}[leftmargin=*]\item Used for the first time a special fiber design that maintains the OAM states while propagation\end{itemize}\vspace{-\topsep}\\
\hline
 \cite{ChenECOC16}&2016&1.1 km&20 Gbit/s&\vspace{-\topsep}\begin{itemize}[leftmargin=*]\item Demonstrated the first bidirectional multiple-OAM communication over a vortex fiber (20 Gbit/s Uplink/Downlink) \end{itemize}\vspace{-\topsep}\\
\hline
 \cite{HuangSREP15}&2015&5 km&40 Gbit/s&\vspace{-\topsep}\begin{itemize}[leftmargin=*]\item Established a communication link using OAM$_{-1,0}$ and OAM$_{+1,0}$  over a conventional graded index FMF at 1550 nm
\item Used a $4\times4$ MIMO DSP to mitigate inter-channel crosstalk
 \end{itemize}\vspace{-\topsep}\\
\hline
\cite{ConventionalMMF1,ConventionalMMF2}&2016&2.6 km&40 Gbit/s&\vspace{-\topsep}\begin{itemize}[leftmargin=*]\item Conducted a MIMO free transmission over a conventional MMF with low crosstalk \end{itemize}\vspace{-\topsep}\\
\hline
\cite{ConventionalMMF3}&2018&8.8 km&120 Gbit/s&\vspace{-\topsep}\begin{itemize}[leftmargin=*]\item Demonstrated a 6 OAM mode (from two mode groups OAM$_{\pm,1}$ and OAM$_{0,1}$) transmission over a conventional MMF
\item Used $2\times2$ and $4\times4$ MIMO DSP to compensate for intra-group channel crosstalk
 \end{itemize}\vspace{-\topsep}\\
\hline
\cite{OAMFMFMMF}&2018&5.6 km&40 Gbit/s&\vspace{-\topsep}\begin{itemize}[leftmargin=*]\item Successfully transmitted two OAM modes, OAM$_{-1,1}$ and OAM$_{0,1}$, over a heterogeneous fiber link (2 km FMF+2.6 km MMF+1 km FMF) without using MIMO DSP at the receiver side\end{itemize}\vspace{-\topsep}\\
\hline
\cite{RingCore}&2018&18 km&8.4 Tbit/s&\vspace{-\topsep}\begin{itemize}[leftmargin=*]\item Combined OAM multiplexing and WDM by encoded information over two OAM modes, OAM$_{+4}$ and OAM$_{+5}$, over 112 wavelength channels\end{itemize} \vspace{-\topsep}\\
\hline
\cite{12ModeOPEX}&2018&1.2 km&10.56 Tbit/s&\vspace{-\topsep}\begin{itemize}[leftmargin=*]\item Transmitted 12 OAM modes with WDM, using 60 25-GHz-spaced channels (with 10 Gbaud signals). 
\item No MIMO processing was  used at the reception.\end{itemize} \vspace{-\topsep}\\
\hline
\end{tabular}
\end{table*} 
\indent Propagating OAM over MCFs has also been suggested in \cite{MCFOAM1,MCFOAM2}. In \cite{MCFOAM1}, Li and Wang proposed a novel design of a 19 ring-core fiber, with each core supporting 22 modes, including 18 OAM states. The authors stated that such fiber might enable the data transmission capacity of Pbit/s, and provide ultra-high spectral efficiency levels.\newline
In \cite{OAMFSOFiber}, Jurados-Nadas et al. presented a hybrid OAM fiber/free space communication scheme, at a wavelength of 850 nm, which offers great potential for propagating OAM over fibers. Their idea was to show that, when multiple superimposed OAM modes after propagating in free space can be coupled to a fiber, without the need for intermediate devices to convert optical signals to electrical signals and then generate optical signals to be conveyed over fiber modes \cite{OAMFSOFiber}.\newline

\subsection{Underwater Communication}
A typical choice for oceanographic data acquisition is to deploy underwater sensors that record data during the monitoring mission, and subsequently recover the information from the sensor's storage unit. However, such an off-line approach cannot deliver real-time information which can be particularly critical in surveillance and disaster prevention. Additionally, there is an increasing use of robotics in underwater missions in order to increase precision and operability. Remotely-operated vehicles (ROVs) and autonomous underwater vehicles (AUVs) are generally used in these types of operation, and an essential part of  is how to communicate with the ROVs or AUVs. 
This communication is usually done using cabled or fiber-based techniques \cite{TiveyRE00}. However, although these approaches provide high speed and reliable transmission of data, their use can be challenging in locations with difficult access and in deep sea. Indeed, in those cases, they will limit the range and maneuverability of the ROVs. 
Wireless communication techniques are much more appropriate for such applications. Wireless transmission under water can be achieved through radio, acoustic, or optical waves. Traditionally, and for more than 50 years, acoustic communication has been used for underwater applications, as it can cover long distances, up to several kilometers \cite{ChitreMTSJ08}. 
The typical frequencies associated with this type of communication are 10 Hz to 1 MHz. However, it is well known that this technology suffers from a very small bandwidth availability \cite{ObergOCConf06}, very low celerity, lack of stealth, and large latencies due to the low propagation speed \cite{PignieriMILCOM08}. As such, data rates using underwater acoustic communication are limited to a few hundreds or thousands of kbit/s. RF waves suffer from high attenuation in sea-water and can propagate over long distances only at extra low frequencies  (between 30 and 300 Hz). This requires large antennas, with high transmission powers, making them unappealing for most practical purposes.\newline
\indent Underwater wireless optical communication (UWOC) has been extensively studied, and recently, high bit rates were reached over tens of meters.
In comparison with the well-mastered acoustic communication techniques, optical wireless communication through ocean water can cover large bandwidth and involve transmitting high data rates exceeding 1 Gbit/s for short and moderate ranges \cite{CochenourIEEEJOE08}. Furthermore, the speed of a lightwave underwater is by five orders of magnitude higher than that of an acoustic wave, and enables efficiently and securely carrying information without suffering from latency. \newline
\indent The blue-green region of the light spectrum, typically between 400 and 500 nm, is found to have potential applications for communication through pure water, as it attenuates the transmitted optical signals very little, in comparison with other light spectrum in the visible/mid-infrared regions \cite{HaleAO73}. With recent advances in solid-state lighting and the development of low-cost and energy-efficient light emitting diodes (LEDs) and laser diodes (LDs), UWOC is more promising than ever, reaching transmission capacities of a few Gbit/s over short and moderate distances \cite{BaidenOCE,NakamuraOPEX15,OubeiOPEX15I,OubeiOPEX15II,ShenOPEX16,XuPTL16,UWOCDiode,Al-HalafiJOCN17,KongOPEX17}.\newline
\indent A possible way to increase the transmission capacity of UWOC link is to use multiple OAM modes, as demonstrated in \cite{BaghdadySPIE15,BaghdadyOE16,BaghdadyIPC,RENSREP16,ZhaoOPEX17} and reviewed in \cite{UWOCWillner}. Baghdady et al. performed experimental demonstration of a 3 Gbit/s underwater communication link, using two 445 nm lasers, and successfully transmitted the signal over a distance of 2.94 m modulated on a simple on-off keying non-return to zero (OOK-NRZ) scheme \cite{BaghdadyOE16}. In \cite{RENSREP16}, Ren et al. reported an aggregated data rate of 40 Gbit/s, using 4 green OAM modes over 1.2 m of turbulent tap water. Each of the four beams carried a 10 Gbit/s signal generated by doubling the frequency of the data signal, at 1064 nm (infrared light). This was achieved using a periodically poled lithium niobate (PPLN) nonlinear crystal, and an integrated dielectric metasurface phase mask, specially designed to imprint the OAM structure on the incoming laser beam, as previously demonstrated in \cite{ArbabiNATNANO15}.
Zhao et al. demonstrated in \cite{ZhaoOPEX17} a single-channel-to-many communication using 4 OAM modes modulated with 8-QAM orthogonal frequency division multiplexing (OFDM) modulation format. The authors also considered the 16-QAM-OFDM and 32-QAM-OFDM schemes to investigate the effect of changing the modulation formats.\newline
Communication involving different propagation media was also proposed \cite{WaterAirWater}. In their experimental studies, Wang et al. showed that water-air-water communication using OAM beams could achieve a bit rate of 1.08 Gbit/s for a single channel \cite{WaterAirWater}. An adaptive feedback-enabled technique has been used to account for distortions caused by change of the water surface.\newline
Non-line-of-sight (NLoS) underwater communication using OAM was further demonstrated. The idea consisted of using the water air interface as a mirror, using the total internal reflection principle, to connect two underwater communicating terminals having a blocked direct LoS path \cite{NLOSOAM}.\newline
Equipments to perform UWOC are commercially available by Sonardyne, STM and Aquatec group companies \cite{Bluecomm,Anglerfish,Aquamodem}. The specifications of the available UWOC solutions are summarized in Table \ref{UWOCSpec}. Including spacial modes will potentially lead to higher performances of those UWOC devices.\newline
\begin{table}[!t]
\renewcommand{\arraystretch}{1.3}
\caption{Specifications of commercially available UWOC systems}
\label{UWOCSpec}
\centering
\begin{tabular}{|l|l|l|l|}
\hline
\rowcolor[gray]{0.8}
 Company&Model &Throughput&Range\\
\hline
\cellcolor[gray]{0.8}&Bluecomm 100&5 Mbit/s&10 m\\                                                                    
\cline{2-4}
\cellcolor[gray]{0.8}Sonardyne \cite{Bluecomm}&Bluecomm 200&12.5 Mbit/s&150 m\\
\cline{2-4}
\cellcolor[gray]{0.8}&Bluecomm 5000&500 Mbit/s&7 m\\
\hline
\cellcolor[gray]{0.8}STM \cite{Anglerfish} &Anglerfish&-&50 m\\
\hline
\cellcolor[gray]{0.8}Aquatec Group \cite{Aquamodem}&Aquamodem Op2L&10.5 KBytes/s&500 m\\
\hline
\end{tabular}
\end{table} 

\subsection{Wireless Indoor Connections}
\begin{figure*}[!t]
\centering
\includegraphics[width=7in]{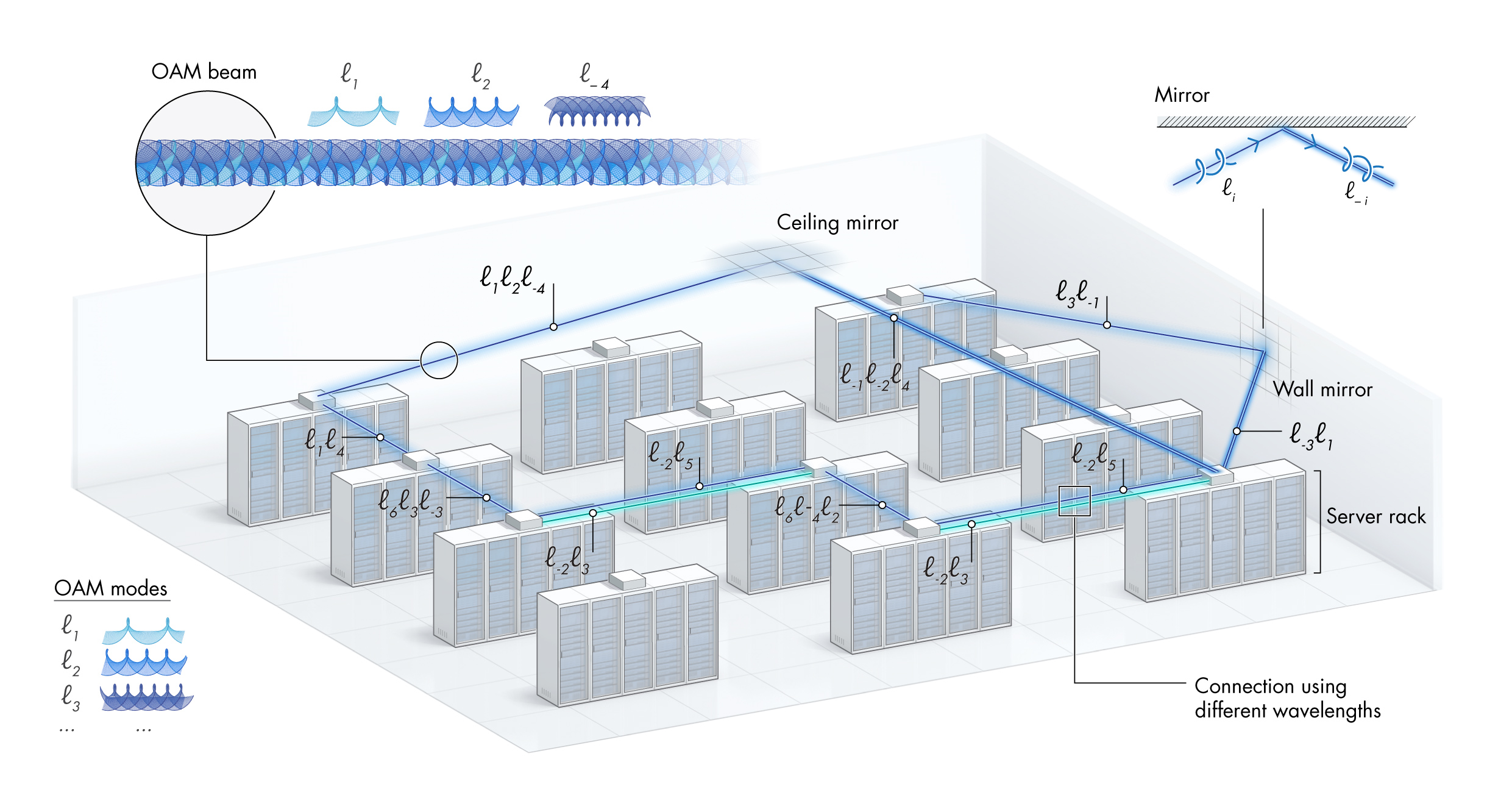}
\caption{Schematic of a potential OAM wireless indoor inter-rack connection in a data center. We note that, when reflected by a mirror, an OAM beam changes the handedness, i.e., it gets the opposite topological charges.  \copyright King Abdullah University of Science and Technology.}
\label{OAMDataCenters}
\end{figure*}
Driven by the dramatic growth in cloud services, data centers have become an essential element of modern communication networks. With an expected annual global IP traffic of 3.3 ZB (1e9 Gigabytes [GB]) in 2021 \cite{CISCO}, high-bit rate communications need to meet the ever-growing demand in data. The installation of optical fibers came at a fundamental cost: density. Energy consumption in data centers is also a major concern. The electricity usage consumed by data centers is estimated to be 1.3 \% of the worldwide electricity usage and for countries such as the United States, the electricity usage of data centers is estimated at almost 2 \% of the entire electricity usage \cite{Electricity1,Electricity2}. \newline
\indent One of the 5G horizons is to establish new optical interconnection techniques, beyond fiber and wire infrastructures, consisting of indoor wireless optical links that offer high transmission capabilities over very short distances. According to Halperin et al.\cite{DCOWC} and Arnon \cite{Arnon}, free space communication can significantly reduce energy consumption, in addition to the high bit rates and low latency. Wireless optical communication links can equally offer a flexible and dynamic reconfiguration of the network. 
OAM multiplexing has been proposed to establish indoor wireless connections for data centers \cite{KupfermanCSNDSP16,Kupferman2017}. The idea is to install a transceiver on top of each rack to communicate with neighbouring transceivers and create adaptive optical wireless network, as illustrated in Fig.~\ref{OAMDataCenters}. Direct links between racks can be implemented, and ceiling as well as reflecting mirrors can be used to steer the OAM beams between two shadowed racks.  With such an OAM-based wireless architecture, it is also possible to generate and transmit the same OAM modes over different wavelengths. \newline
\indent Engineering guidelines for the design of an OAM multiplexing-based communication system for data centers applications is provided in \cite{OAMDataCenterDesign}. An algorithm for encoding and decoding information, using the patterns of the OAM modes that can potentially be suitable for data center applications, is proposed in \cite{KupfermanJO18}.

\subsection{On-chip Photonic Circuits}
\begin{figure}[!t]
\centering
\includegraphics[width=3in]{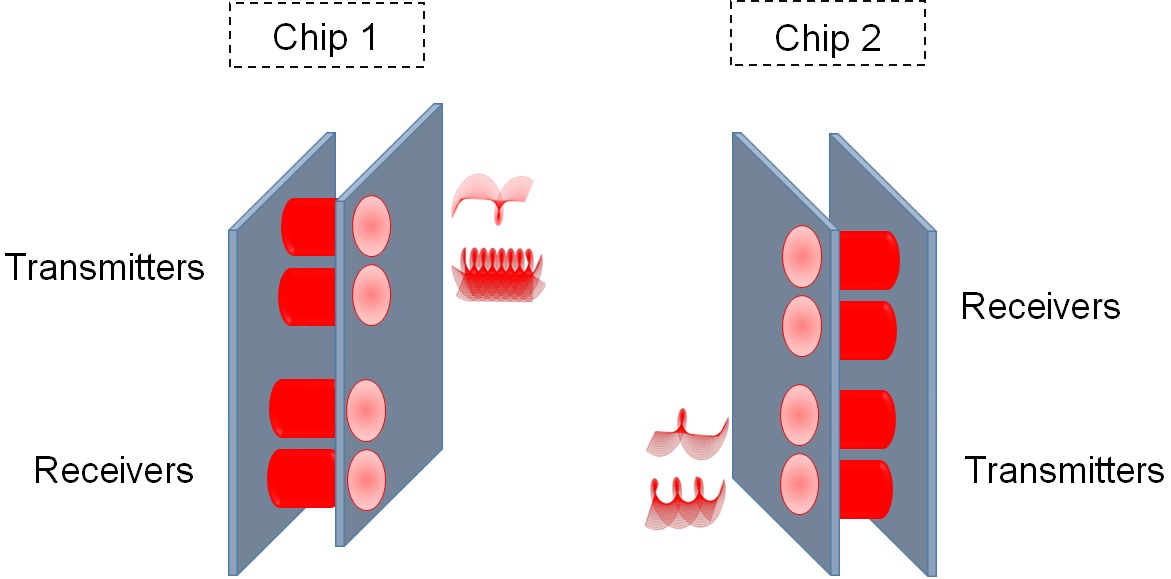}
\caption{A possible chip-to-chip OAM intectonnection structure.}
\label{OAMChip}
\end{figure}

Improving the performance of on-chip circuits is key to production of future data servers and faster computers. The on-chip data bandwidth in an electronic integrated circuit is dependent on the metal interconnect being used, which can significantly affect the system performance \cite{OnChip1}. An alternative and reliable technology is the technology using integrated photonic circuits that are able to maintain a high data transmission bandwidth and reduce the consumption of energy \cite{OnChip1}. Advances in complementary metal-oxide-semiconductor (CMOS) technology have made it possible to fabricate on-chip photonic circuits that are energy-efficient and have a high data bandwidth \cite{OnChip2,OnChip3}.\newline
\indent Al$\grave{u}$ and Engheta proposed using wireless optical interconnects, at the microscale and nanoscale levels, to achieve higher performances for on-chip communications \cite{OnChip4}. In their study, the authors only considered plane and spherical waves, which inspired Zhang and his co-workers, two years later, to investigate the potential of a multimode wireless optical on-chip scenario \cite{ZhangOPEX12}.
Ultimately, the authors proposed a full solution to encode and decode data over OAM beams, with a topological charge ranging from $\ell=-4$ to 3 for optical wireless on-chip interconnects. An OAM on-chip structure is schematized in Fig.~\ref{OAMChip}.\newline
 Liu et al. performed comparative studies showing that OAM-based optical wireless interconnect has the highest power efficiency, compared with a PHY-based electrical interconnect, an interposer-based microbump, and a dense-WDM interconnect \cite{LiuIEEEDT}. \newline 
\indent In 2016, Wang et al. presented an on-chip generation technique using 4 OAM modes, detailing their experimental studies using a dielectric metasurface based on a silicon platform \cite{WangECOC16}. A digital holography-based approach was later proposed by Li et al. for the multiple encoding and decoding of OAMs, which can potentially be applied to the production of OAM on-chip interconnects, on a large scale \cite{LiOPEX17}. 
The only limitation for the production of these new types of interconnects on nanoscale chips comes from the fact that each pixel of a liquid crystal display (LCD) of a SLM has a pitch of a few $\mathrm{\mu}$ms (8 $\mathrm{\mu m}$ for SLMs of Holoeye \cite{Holoeye}) which is not enough to manipulate small-sized beams. Luckily, newly designed small-footprint OAM mode emitters \cite{OAMGeneration4,OAMDetectionPCircuit,3DOAMDetector}  and detectors are proposed (see details in section \ref{Sec:GenDet}).
A tunable on-chip OAM and WDM de-multiplexer based on microelectromechanical systems (MEMS) technology was described in \cite{MEMSChip}. A bit rate of 64 Gbit/s over a 4-OAM chip-to-chip transmission using a pair of silicon photonic multiplexer/demultiplexer was reported in \cite{NewOnChip}. \newline
\indent Similar to free space and underwater communication, multicasting function can be also implemented on microscale chips, as reported by Du and Wang who designed a V-shaped array to generate a 4-fold OAM from a Gaussian beam for a low-crosstalk one-onto-many channel distributions \cite{Multicastingonchip}. \newline
\subsection{Optical Switching and Routing}
\begin{figure}[!t]
\centering
\includegraphics[width=3.5in]{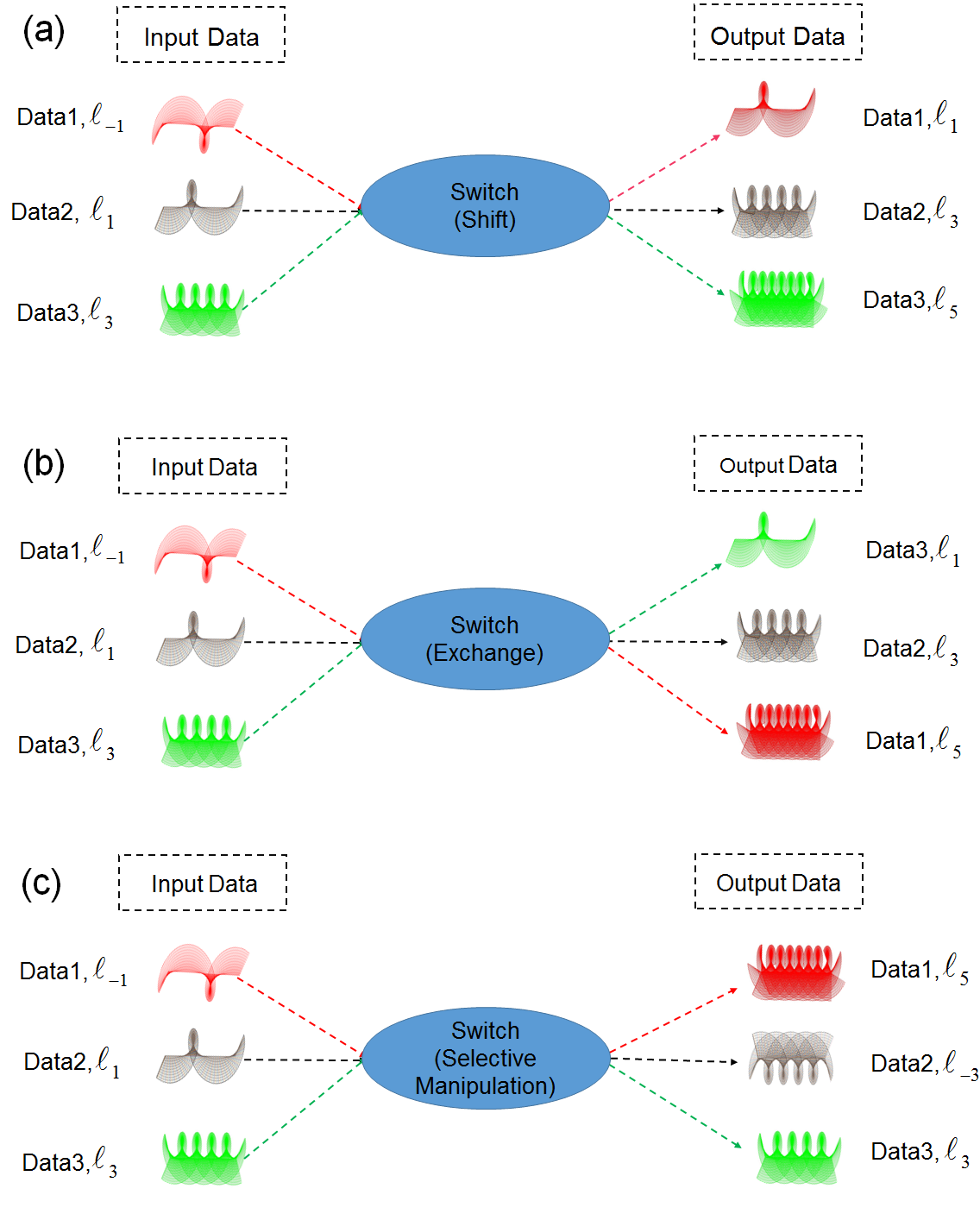}
\caption{Schematic showing different switch configurations: (a) OAM index shift, (b) OAM index exchange and (c) OAM index selective manipulation. }
\label{OAMSwitch}
\end{figure}

Electronic switches have been adopted in optical communication networks for a long period of time \cite{WDMSwitch1}. In every switching node, optical signals are converted to electrical ones, buffered electronically, and then re-routed to their next node after being reconverted to the optical form. However, this technology consumes a great amount of power and bandwidth \cite{WDMSwitch1}. Optical switching, in particular, WDM switching, provides a good alternative to electronic switching, as they enable routing signals without the need for their conversion to electrical signals, thus significantly increase the network throughput and the decrease of switching time \cite{WDMSwitch2}.
Similarly to wavelength switching, it is possible to use space as an additional degree of freedom to perform switching and routing functions \cite{SDMSwitch}.\newline
One potential spatial mode set for optical space switching is OAM as reported in \cite{Routing1,Routing2,Functions}. It is possible to perform various switching functions using OAM modes including:
\begin{itemize}[leftmargin=0.5cm]   
\item Shift function that consists of shifting all the OAM beams by the same index step $k$ while keeping the same data carried by each mode. The topological charge $\ell$ of a particular beam becomes $\ell+k$. As presented in Fig.~\ref{OAMSwitch}(a), the OAM beams with charges  ($\ell=-1$, $\ell=1$, $\ell=3$), respectively, carrying data streams (Data$_{1}$, Data$_{2}$, Data$_{3}$) are shifted to ($\ell=1$, $\ell=3$, $\ell=5$) after the shift function with $k=2$.
\item Exchange function which means reversing the order of the OAM charges; data carried on the highest OAM charge becomes carried on the lowest OAM charge and vice versa. As depicted in Fig.~\ref{OAMSwitch}(b), data information (Data$_{1}$, Data$_{2}$, Data$_{3}$) carried by ($\ell=-1$, $\ell=1$, $\ell=3$), respectively, are carried by  ($\ell=5$, $\ell=3$, $\ell=1$) after the Exchange function. The Exchange function was experimentally implemented between two OAM beams, carrying 100 Gbit/s DQPSK data signals \cite{WangNATPHO12}.
\item Selective manipulation function is different from the Shift and Exchange functions and can be applied on one or multiple OAM states without changing the others. The idea is to switch one beam with a charge $\ell$ to unoccupied charge $q$ without affecting the other OAM beams. As shown in Fig.~\ref{OAMSwitch}(c), $\ell=-1$ (Data$_{1}$) is switched to $\ell=5$ beam and $\ell=1$ (Data$_{2}$) is switched to $\ell=-3$ while $\ell=3$ (Data$_{3}$) remains unaffected. 
\end{itemize}
 OAM switching, associated with polarization control was further demonstrated on 100 Gbit/s QPSK signals \cite{RoutingOAMPOL}.\newline
Willner and his co-workers demonstrated a $2\times2$ OAM reconfigurable switch \cite{Routing3,Routing4} using SLMs for the OAM down and up-conversions. Designing an $N\times N$ switch may require the use of up to $2N+1$ spatial light modulators \cite{Routing2}. 
As we previously mentioned in section \ref{Sec:GenDet}, liquid crystal SLMs are relatively slow to respond limiting the switching rates to the order of 1 kHz in the best case scenario (Boulder Nonlinear Systems SLMs). In \cite{Routing5}, Scaffardi et al. demonstrated OAM based $2\times2$ reconfigurable switch faster than the one reported in \cite{Routing3,Routing4} and with less then 1 dB back to back power penalty. The switch is suitable for signals modulated up to 20 Gbaud and the main contribution of this work is that, instead of using an SLM for switching between the OAM states, the switching component is implemented on a silicon-on-insulator chip exploiting microring, allowing fast change in the order of nano-seconds.  
By designing a novel gratings implemented on digital mircromirror device, authors in \cite{Routing6} achieved a fast data exchange and multicasting in 49 OAM channels with an aggregated transmission capacity of 1.37 Tbit/s, at a switching rate of 6.9 $\mathrm{\mu s}$. In \cite{InterconnectionOAM}, a three layer OAM, polarization and wavelength switch architecture-based integrated concentric OAM emitters/modulators are demonstrated.
\subsection{Radio Frequency Communication}
\begin{figure}[!t]
\centering
\includegraphics[width=3in]{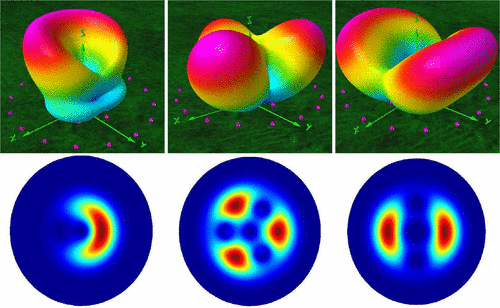}
\caption{Beams obtained by superimposing two different OAM states. The upper three panels show the radiation patterns for the antenna array, and the lower three panels show the corresponding intensity patterns, head on, calculated for Laguerre-Gaussian beams. The leftmost are for $\ell_{1} = 1$ and $\ell_{2}= 2$, the middle ones are for $\ell_{1} = 1$ and $\ell_{2} = 4$, and the rightmost are for $\ell_{1} = 2$ and $\ell_{2} = 4$. Notice the good agreement between the patterns obtained with the antenna array model and the paraxial LG beam model \cite{ThidePRL07}.}
\label{OAMRF}
\end{figure}

OAM modes have been further evidenced in RF communication, where multiple OAM beams can share the same frequency and provide a solution to the spectrum scarcity. Thid\'e et al. were the first to significantly highlight the potential of using OAM modes for radio communication \cite{ThidePRL07}. They numerically showed that vectorial antenna arrays are able to generate radio beams having spin and angular characteristics similar to LG beams in optics at a frequency lower than 1 GHz. Fig.~\ref{OAMRF} shows the radiation patterns as well as the intensity patterns of superposition of twisted radio beams. Through a system simulation study of radio OAM, Mohammadi et al. addressed the generation and measurements of OAM beams at radio frequencies by designing a circular antenna array capable of generating and receiving radio OAMs \cite{MohammadiIEEETAP10}. Tamburini et al. successfully carried out the first experimental test of outdoor transmission using multiple OAM beams over the same radio frequency in Piazza San Marco in Venice \cite{TamburiniNJP12}.
A video transmission of 4 Gbit/s over 60 GHz radio OAM was reported in \cite{OAMVideo}. In \cite{YanNATCOM14}, Yan et al. presented a new 32 Gbit/s millimeter-wave communication system with a spectral efficiency of 16 bit/s/Hz was demonstrated, using 4 radio OAM beams transmitted over a 2.5 m link. In a separate study, Yao et al. reported the use of commercially available impulse radios and SPPs, to realize a two-OAM 3 Gbit/s indoor transmission over a 2 m link, with a BER of $9.65\times10^{-10}$ and a crosstalk of -15 dB for both beams \cite{CommercialAntenna}.\newline
\indent Djordjevic proposed a multidimensional OAM-based coded modulation scheme for future generation wireless communication systems \cite{Djordjevic}. The author claimed that the proposed modulation technique enables beyond 1Tb/s RF communications ensuring a high physical layer security. A numerical study by Zhang et al. highlighted the potential for using an OAM MIMO-based communication scheme to further increase the capacity of OAM multiplexing systems \cite{OAMMiMo}. A 16 Gbit/s dual-OAM beam mm-wave transmission combined with a conventional $2\times2$ MIMO scheme, at a carrier frequency of 28 GHz, was presented in \cite{RenIEEETWS}. Simulation results reported in \cite{WangACCESS17} showed that OAM-MIMO outperformed conventional MIMO wireless communication. Combining massive MIMO with OAM is expected to maximize the spectral efficiency of traditional massive MIMO mm-wave radio communications, as shown in \cite{OAMmassiveMIMO}. Ge et al.  proposed in \cite{SMOAM} an OAM spatial modulation (SM) millimeter wave communication scheme where for an $M$ transmission antenna configuration, and at a given time slot, only one antenna is set to transmit OAM beams. The authors reported more than 200\% improvement of the the maximum energy efficiency compared to the OAM-MIMO configuration and the proposed scheme showed more resilience against path-loss attenuation. However, the impact of antenna misalignment was not investigated. For more details on spatial modulation, readers may refer to \cite{SpatialModulation}.\newline
\indent A recent study revealed that using the RF mode dimension could offer a solution to jamming problem in wireless communication, where the authors proposed the use of the orthogonality of OAM modes to provide mode hopping (MH) with a narrowband frequency which can be further combined also with the conventional frequency hopping (FH) \cite{Jamming}. Using radio OAM beams for radar applications was further studied in \cite{OAMRadar1,OAMRadar2}. Authors in \cite{OAMRadar2}, reported an experimental demonstration of a high resolution synthetic aperture radar (SAR) imaging using OAM waves and found that due to their wider width, OAM beams perform better than plane waves in SAR applications.
\subsection{Acoustic Communication}
OAM multiplexing is not restricted anymore to electromagnetic (EM) waves-based communications, and now includes acoustic wave transmission. It has been proven that the use of OAM structured acoustic waves improves the capacity of underwater acoustic communication. Using 8 OAM states, the spectral efficiency of the communication system reaches $8.0\pm0.4$ bit/s/Hz, at an SNR of 20 dB and a BER below $10^{-6}$ \cite{PNAS17C1}. The reported measured crosstalk is lower than -8.54 dB. The main objective of the study is to highlight the potential of the use of multiple acoustic vortex beams to increase the transmission capacity of existing long range acoustic wave communication systems.\newline
\section{OAM Challenges}
The real implementation of OAM communication systems in any of the applications discussed here requires overcoming various challenges that depend both on the propagation medium and the adopted mode generation/detection techniques. Here, we review the main challenges related to OAM communication applications.
\subsection{Free Space Propagation Effects}
\label{FSOSubSection}
When propagating through the air, optical signals are subject to three major effects: absorption, scattering and turbulence.
\begin{itemize}[leftmargin=0.5cm]   
\item Absorption and scattering are caused by the interaction of laser beams propagating in free space with various molecules of gas, including NO$_{2}$ and O$_{2}$, and other small particles in the earth's atmosphere that lead to power losses; indeed according to Beer's law \cite{BeerLaw}, the power of a beam decreases as it propagates, following an exponential function ruled by an attenuation coefficient. These phenomenea are mainly related to the weather conditions including fog, rain and snow \cite{KedarIEEECM04}.
\item Turbulence is related to temperature and pressure fluctuations in the atmosphere resulting in a random behavior in the atmospheric refractive index \cite{Andrews}. 
The propagation of an optical beam in turbulent atmosphere leads to wavefront distortion in addition to beam spread and wandering.\newline
\end{itemize}
For 4 co-propagating OAM beams each carrying a 20 Gbit/s QPSK signal, the effect of ballistic scattering and diffuse scattering was investigated in \cite{Scattering}. It was been found that ballistic scattering leads to power loss, whereas diffuse scattering induce inter-OAM channels crosstalk.
The effect of atmospheric turbulence on the OAM states was investigated thoroughly in \cite{Anguita,FSOTurbulenceOAM1,FSOTurbulenceOAM2}. Anguita et al. numerically investigated the effect of different atmospheric turbulence regimes on the simultaneous propagation of multiple OAM beams in free space \cite{Anguita}.
By randomly varying the phase aberration following Kolmogorov's turbulence model, Rodenburg et al. investigated the crosstalk between 11 distinct OAM modes and found that the turbulence significantly degraded the mode purity, independently of the OAM charge \cite{FSOTurbulenceOAM1}. The effect of power spreading between neighbouring modes on the system performance is shown in \cite{FSOTurbulenceOAM2}. \newline
 Lavery et al. experimentally studied the effect of turbulent atmosphere above the city of Erlangen (Germany) on the propagation of OAM modes, over a 1.6 km outdoor free space link \cite{TurbulenceGermany}. Turbulence induced beam wandering causes crosstalk for OAM modes, which, at the system level, deteriorates the BER performance, and increases the power penalty. This phenomenon was verified by several experimental studies investigating the propagation of OAM beams \cite{Ren400G}. Ren et al. conducted an outdoor 4 OAM experimental transmission with an aggregate capacity of 400 Gbit/s over a 120 m turbulent free space link, in order to investigate the influence of atmospheric inhomogeneities on the OAM received powers, the inter-modal crosstalk, and the system power penalties \cite{Ren400G}. The effect of pointing errors was further investigated in the same paper. 
Additional deployment challenges of OAM communication over turbulent channels were discussed \cite{WillnerJOP16,HuangIEEETWC18,MingLi1}.\newline
\indent In laboratories, atmospheric turbulence conditions have been simulated in laboratories, using SLMs \cite{TurbulenceSimulationSLM}, heated pipes \cite{TurbulenceSimulationPipe}, static diffractive plates \cite{TrichiliOL16} and rotating phase plates \cite{RotatingPlate}, in order to analyze its effects on OAM carrying beams. The most common model for the aforementioned laboratory-emulated turbulence is the Kolmogorov model. 
Following Kolmogorov's model, on the presence of atmospheric turbulence, the electric field of the OAM mode, $E_{\ell}$, becomes:
\begin{equation}
U(r,\phi,z)=E_{\ell}(r,\phi,z)\exp(i\pi \sum_{n}^{\infty}\sum_{m=0}^{n} c_{n,m}Z_{n}^{m}(r,\phi)),
\end{equation}
where $Z_{n}^{m}$ is the Zernike polynomial of radial order $n$ and azimuthal order $m$, and $c_{n,m}$ is the aberration strength in the form of a simple weighting factor \cite{zernike}. The turbulence aberration strength can also be characterized by the Strehl ratio (SR), as explained in \cite{SR}. The SR decreases with the atmospheric distortions, with a value of 1 corresponding to a turbulence-free channel. However, there are evidences that the Kolmogorov's turbulence model has limitations in predicting the lower layers of the atmosphere and also when the atmosphere is stable \cite{Non-Kolmogorov}. \newline 
\indent In the case of Gaussian plane waves, the formulas to determine the bit error rate, outage probability and the channel capacity are provided in the literature for a wide range of turbulence models, including lognormal for weak turbulence \cite{LogNormal}, K-distribution for strong turbulence \cite{KDistribution}, negative exponential for very strong turbulence \cite{NegativeExpo}, Gamma-Gamma \cite{GammaGamma} from weak to strong turbulence, and the recently introduced Malaga distribution \cite{Malaga}. However, such a theoretical framework is still missing for the case of the OAM, in order to show whether the channel key performances are $\ell$ dependent or not.\newline
We should note that free space turbulence may also occur in an indoor environment due to ventilation from cooling systems in servers which should be taken into account, when establishing wireless indoor interconnections.\newline 
\indent In addition to being distorted, when propagating through turbulence media, OAM beams tend to diverge and the divergence increases with $\lvert\ell\lvert$ \cite{DivergencePadgett}. This could be a serious issue for limited size optical receiving apertures when some of the power carried by the beam is lost, due to beam divergence. The impact of the receiving aperture size and the misalignment between the transmitter and the receiver on the performance of an OAM FSO communication link was carefully investigated in \cite{OpticaFSO}. Taking into account the divergence-induced power loss, Xie et al. of  studied the effect of misalignment caused by lateral displacements between the transmitting and the receiving terminals \cite{OpticaFSO}. They also investigated the effects of an angular error in the receiver on the system power penalty, channel crosstalk and power loss.
\subsection{Fiber Propagation Effects}
Linear and nonlinear effects are the main challenges for OAM propagation in fibers.
\subsubsection{Fiber Linearity}
One of the main linear fiber effects is the attenuation due to material absorption. Attenuation levels for fibers carrying OAMs, including FMFs, MMFs and vortex fibers, are usually small and comparable with standard SMFs.
Another phenomenon that occurs in optical fibers is dispersion, usually referred to 'chromatic dispersion'. Chromatic dispersion is the result of the fact that the phase velocity and the group velocity of propagating light beam in a fiber depends on the optical frequency. Dispersion coefficients can be obtained through a Taylor expansion of the propagation constant of the lightwave at an angular frequency $\omega$, around a central frequency $\omega_{0}$:
\begin{equation}
\beta(\omega)=\beta_{0}+\frac{\partial\beta}{\partial\omega}(\omega-\omega_{0})+\frac{\partial^2\beta}{\partial\omega^2}(\omega-\omega_{0})^2+\frac{\partial^3\beta}{\partial\omega^3}(\omega-\omega_{0})^3+...,
\end{equation}
$\beta_{0}$ accounts for a phase shift, and $\beta_{1}=\frac{\partial\beta}{\partial\omega}$ denotes the group delay per unit length. The second order term $\beta_{2}=\frac{\partial^2\beta}{\partial\omega^2}$ is known as the group velocity dispersion (GVD), whereas $\beta_{3}=\frac{\partial^3\beta}{\partial\omega^3}$ represents the third order dispersion (TOD).
Two other types of dispersion may occur in fibers: the intermodal dispersion and polarization mode dispersion (PMD). The former is a phenomenon in which group velocity depends not only on the frequency, but also on the polarization state, whereas in the latter, the phase velocity depends on the optical mode. The strength of the PMD, or the time delay between two polarization states, is quantified by the differential group delay (DGD) whereas the modal dispersion is quantified by the  differential mode group delay (DMGD).
Additional linear effects can be possibly created due to core ellipticity, fiber twisting and fiber bending.\newline
\indent In \cite{OAMLEffects}, mode properties were theoretically analyzed, including the walk-off and the tolerance to fiber core ellipticity. The authors of \cite{WangOE16} studied the influence of the differential mode delay of two modes over 50 km-long graded-index FMF. In \cite{NejadJLT17}, the effect of DMGD is studied. Wang et al. presented a measurement technique to calculate the coupling between $+\ell$ and $-\ell$ OAM states \cite{OAMPMD}. Referring to single mode fibers, the authors described this effect as a OAM polarization mode dispersion. The first investigation of chromatic dispersion (CD) of OAM modes over an optical fiber was carried out by Yan et al.. In their study, the authors provided a characterization of the CD properties of 3 OAM modes in a ring-core fiber \cite{OAMCDispersion}.
\subsubsection{Fiber Nonlinear Effects}
Different types of non-linearities can occur in optical fibers. Several effects are related to Kerr nonlinearity that arises from the third order susceptibility $\chi_{3}$; this effect can be viewed as a change in the refractive index of a materiel, in response to an electric field. Kerr effects mainly include \cite{Nonlinear}:
\begin{itemize}[leftmargin=0.5cm]   
\item Self-phase modulation (SPM) due to the nonlinear phase delay occurring during the launch of  a high optical intensity $I$ in an optical fiber with a nonlinear phase delay that has the same temporal shape as the optical intensity. This can be described as a nonlinear change in the refractive index of $\Delta n=n_{2}I$, and it results in a phase delay of the optical beam propagating through the medium. $n_{2}$ is the nonlinear refractive index. The phase shift in a fiber length $L$, when dealing with optical power, can be quantified as follows:
\begin{equation}
\phi= n_{2}\frac{2\pi}{\lambda}\frac{P}{A_{eff}}L = \gamma P L,
\end{equation}
where $A_{eff}$ is the effective mode area and $\lambda$ denotes the wavelength of the light signal. The nonlinear coefficient $\gamma$ mostly depends on the fiber properties and less on the wavelength.
\item Cross-phase modulation (XPM) that occurs due to the refractive index being seen by an optical signal in a nonlinear medium depends not only on the intensity of that beam but also on the intensity of another co-propagating beam having a different mode, direction, wavelength or polarization state. The optical intensity of beam $\kappa I^{(2)}$ produces a change of the refractive index of $\Delta n^{(1)}=n_{2}\kappa I^{(2)}$ for beam having intensity $I^{(1)}$. This effect can cause significant crosstalk between communication channels in optical fibers.
\item Four wave mixing (FWM) that occurs when two light signals of different optical frequencies $\omega_{1}$ and $\omega_{2}$ interact in a fiber. FWM generates new signals at frequencies $\omega_{3} = 2\omega_{1}-\omega_{2}$ and $\omega_{4} = 2\omega_{2}- \omega_{1}$.
  \end{itemize}
 Additional nonlinear effects are related to the phonons vibrations namely stimulated Raman scattering (SRS) and stimulated Brillouin scattering (SBS) \cite{Nonlinear}.
\begin{itemize}[leftmargin=0.5cm]  
\item Raman Scattering is a nonlinear process that involves optical phonons meaning high-frequency lattice vibrations in the glass. In the context of Raman active medium, when two light signals having the same polarization direction but at different wavelengths, the signal with longer wavelength can be optically amplified at the cost of the shorter wavelength signal. The evolution of the intensity of the signal, along the propagation direction $z$, is proportional to the existing signal intensity and to the pump intensity.
\item The principle of stimulated Brillouin scattering is very similar to Raman scattering, but instead of involving optical phonons, it uses acoustic phonons of the materiel of the fiber in which the beam propagates. In silica fibers, the back-scattered light has a Brillouin frequency shift of about 11 GHz. SBS can become a dominant effect if the pump power exceeds a threshold value. 
  \end{itemize}
Note that nonlinear effects are detrimental for long distance transmission, in SMFs as well as MMFs \cite{FWMWDM1,FWMWDM2}. However, some effects such as the FWM could be harnessed to perform mode amplification and wavelength conversion applications \cite{FWMFMF1,FWMFMF2}. \newline
In MDM-based communication systems, different types of nonlinearities have been investigated \cite{Nonlinear1,Nonlinear2,Nonlinear3,Nonlinear4}. However, nonlinearity for OAM modes is still a largely unexplored field.
\subsection{Underwater Propagation Effects}
Optical communication through ocean water can offer large bandwidth compared to the acoustic communication techniques. However, in many respects, this technology is still in its infancy and requires further research efforts to overcome several major natural and technical issues before being deployed on a large scale. In particular, quantifying and compensating underwater propagation effects are two crucial steps towards setting up underwater communication links. While propagating in the water, the intensity of a light signal $I$, whether it carries an OAM or not, decays exponentially and approximately following Beer's law, and is expressed as follows:
\begin{equation}
I=I_{0}\exp(-c(\lambda)z),
\end{equation}
where $I_{0}$ is the light intensity at $z=0$, $c(\lambda)$ denotes the attenuation coefficient measured in m$^{-1}$ and depends on the wavelength $\lambda$. The attenuation $c$ is obtained by summing two inherent optical properties $\alpha(\lambda)$ and $b(\lambda)$ which represent two different phenomena, the absorption and the scattering. In addition to the attenuation due to water absorption and scattering, performance of such a communication depends on the underwater environment which is still a major concern. Underwater turbulence affects light propagation due to the changes in refractive index associated with fluctuations in temperature and salinity. UWOC can be also subject to air bubbles that could be mainly produced by breaking surface waves, as highlighted in \cite{Bubbles}. In the case of a single mode UWOC, the effect of temperature was recently modeled and an excellent agreement between experimental data collected for weak thermal-gradient induced turbulence and the generalized Gamma distribution (GGD) is obtained \cite{UWTurbulence1}. Salinity induced turbulence was studied and a Weibull model that provided a perfect agreement with measured data is proposed to model channel fluctuations for a wide range of conditions \cite{UWTurbulence2}. Effect of air bubbles on the underwater propagation was investigated and some link design considerations in a bubbly channel are provided in \cite{UWTurbulence3}. Authors in \cite{UWTurbulence4} found that fluctuations due to air bubbles can be characterized by an Exponential-Gamma (EG) distribution.
\newline
\indent The impact of underwater propagation effects on OAM modes has been theoretically investigated in \cite{ChengAO16,WangIEEEPJ7}. Analytical formulas to determine the OAM probability density and inter-channel crosstalk in weak turbulent ocean are provided in \cite{ChengAO16}. Under a weak oceanic turbulence regime, Wang et al. derived analytical formulas for the channel capacity and symbol error rate (SER) for OAM channels \cite{WangIEEEPJ7}. In Table \ref{UnderwaterConditions}, we summarizes the main experimental studies investigating the impact of  underwater conditions on OAM modes covering mainly scattering, thermally induced turbulence, water turbidity and air-bubbles \cite{CochenourAO16,RENSREP16,ViolaCLEO16,MorganJOpt16,ZhaoOL17}.
\begin{table*}[!t]
\renewcommand{\arraystretch}{1.3}
\caption{Literature on the experimental investigation of turbulence conditions for underwater transmission using OAM}
\label{UnderwaterConditions}
\centering
\begin{tabular}{|>{\columncolor[gray]{0.8}}l|p{0.2\textwidth}|l|p{0.4\textwidth}|}
\hline
\rowcolor[gray]{0.8}
Reference & Investigated Propagation Effect &Link Length&Main Results and Contributions\\
\hline
\cite{CochenourAO16}&Scattering&1 m&\vspace{-\topsep}\begin{itemize}[leftmargin=*]\item Investigated the effect of scattering on 3 OAM states ($\ell=0$, $\ell=8$ and $\ell=16$) using a commercial anti-acid, Maalox.
\item Revealed dependency between scattering and OAM states (received power increases with the OAM order)
\item Consistent results with \cite{AlfanoOL16} that used a sample cell to study the effect of propagation through a scattering media
\end{itemize}\\
\hline
\cite{RENSREP16}&Thermal gradient, scattering and water-current &1.2 m&\vspace{-\topsep}\begin{itemize}[leftmargin=*]\item Investigated modes $\ell=0$, $\ell=1$ and $\ell=3$ 
\item Thermal-gradient induced turbulence causes OAM beam wandering displacement as well as intermodal crosstalk
\item Scattering leads to significant OAM power loss
\item Water current induces slight beam displacement and wandering
\end{itemize}\\
\hline
\cite{ViolaCLEO16}&Flow similar to oceanic conditions &2.5 m&\vspace{-\topsep}\begin{itemize}[leftmargin=*]\item Involved modes in investigations: $\ell=1$, $\ell=2$ and $\ell=3$
\item Water flow creates non-uniformity on the intensity profiles of the received OAM modes
\item Inter-channel crosstalk.
\end{itemize}\\
\hline
\cite{MorganJOpt16} & Water turbidity similar to oceanic conditions &2 m&\vspace{-\topsep}\begin{itemize}[leftmargin=*]\item Generated a set of petal modes $\ell=(1,-2)$, $\ell=(1,-4)$ and $\ell=(2,-4)$
\item Slight reduction of the spatial and temporal correlation of the OAM mode superposition 
\end{itemize}\\
\hline
\cite{ZhaoOL17} & Bubbles &2 m&\vspace{-\topsep}\begin{itemize}[leftmargin=*]\item Bubbles partially block the OAM modes and induce intensity profile degradation at the reception
\item Severe BER degradation of a signal carried over the $\ell=3$  OAM beam
\item OAM modes are outperformed by Bessel modes under air bubbles induced turbulence
\end{itemize}\\
\hline
\end{tabular}
\end{table*}
We note that the ranges of all the presented investigations from the literature are  limited, which is not consistent with real-life applications requiring propagation distances up to tens of meters.
\subsection{Inter-modal Crosstalk}
Due to disturbance, OAM modes could exchange power while propagating and also power from one mode may generate a neighbouring OAM mode. Such an effect is known as inter-modal crosstalk which is a major limit for implementation of any communication employing OAM modes. As a result to inter-modal crosstalk, there may not be enough power to recover the signal.
Intermodal-crosstalk should be mainly minimized in chip-to-chip interconnects \cite{YuOPEX15}.
{As a result to inter-modal crosstalk, there might not be enough power to recover the signal.}
\subsection{RF Challenges}
\begin{figure}[!t]
\centering
\includegraphics[width=3in]{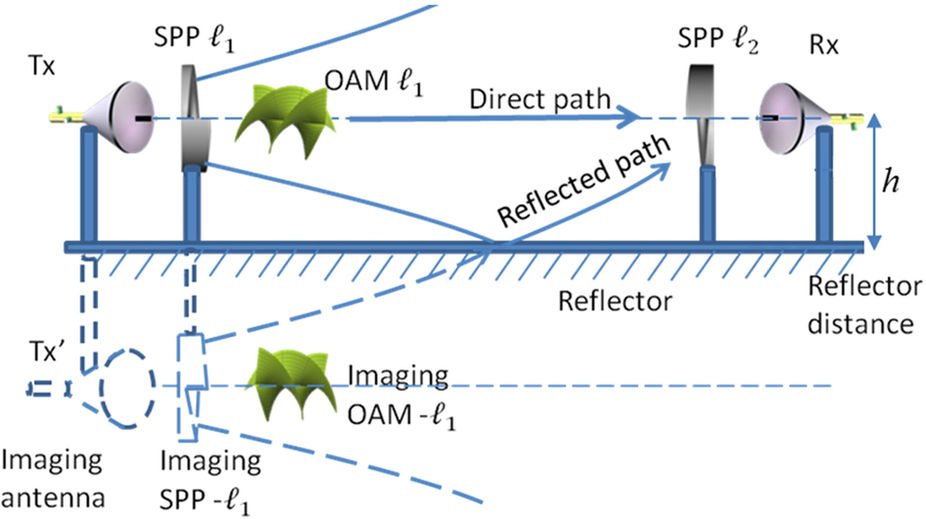}
\caption{Multipath effects of an OAM channel caused by specular reflection from a parallel ideal reflector \cite{Multipath}.}
\label{Multipath}
\end{figure}

 Edfors and Johansson, \cite{OAMRadio} pointed out that bringing OAM to RF can only deliver high performance as well as low complexity, for short distance communications, if OAM states beamforming is performed by a discrete Fourier transform (DFT); they concluded that OAM radio was not optimal for wireless communication. Another study based on theoretical investigations showed that spatial mode multiplexing in RF offered no significant gain, compared to conventional single mode MIMO systems \cite{SDMOAMRadio}. Two side effects are considered to be the main concerns using RF OAM. Those are: 
 \begin{itemize}[leftmargin=0.5cm]
\item Divergence: As in free space, RF OAM beams instinctively diverge when increasing the distance between the transmitter and the receiver \cite{Divergence1,Divergence2}. A schematic illustrating the radio OAM beam divergence is shown in Fig.~\ref{Divergence}. \newline 
\begin{figure}[!t]
\centering
\includegraphics[width=3in]{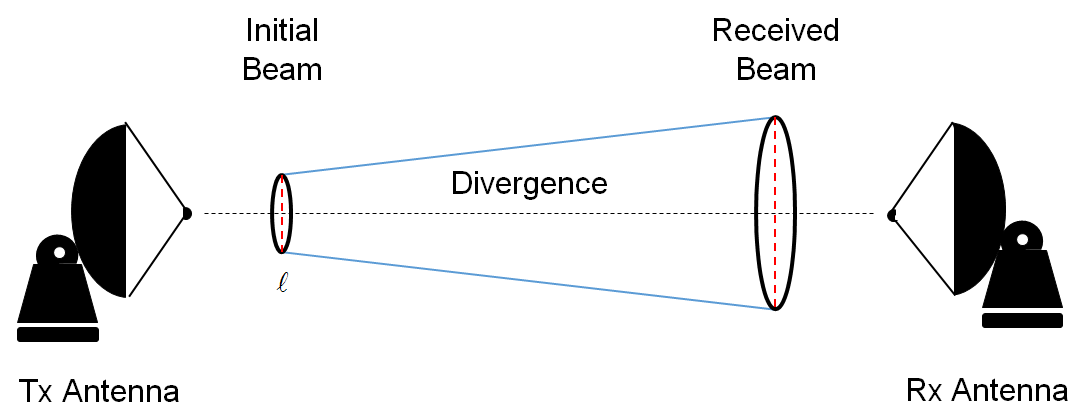}
\caption{Divergence of an OAM radio beam of $\ell$ charge while propagating from a transmitting antenna (Tx) to a receiving antenna (Rx). The divergence scales with $\lvert\ell\lvert$.}\
\label{Divergence}
\end{figure}
\item Multipath: This occurs when a single signal is sent by a transmitting antenna and the receiving antenna receives multiple signals due to reflections, for example from surfaces, water or buildings. A schematic illustrating the effect of multipath on RF OAM mode transmission is shown in Fig.~\ref{Multipath}. The receiving antenna collects not only the initially sent OAM mode with $\ell$ charge, but also a reflected signal with a $-\ell$ charge. 
\end{itemize}
The power received by the receiver decays significantly as a function of the distance, causing inter-channel crosstalk; beams with high $\ell$ indices are more sensitive to power loss, as shown by analytical investigation reported in \cite{Divergence2}. Detecting only a small portion of the radio beam will considerably deteriorate the communication performances and increase the system penalty, which makes RF OAM unsuitable for broadcasting applications \cite{QuestionJO}.\newline
The multipath effect has been studied for OAM radio links as reported in \cite{Multipath}. An analysis, at 28 GHz, of the multipath effect induced by a shiny surface parallel to the communication path revealed that the reflection causes both intra and inter-channel crosstalk meaning that the reflected energy due to multipath is coupled not only to a channel with the same $\ell$, but also to channels with different $\ell$ charges, and leads to distortion of the OAM's wavefront \cite{Multipath}. Beams with larger $\ell$ are believed to be more sensitive to multipath.\newline
Through numerical calculations, Nguyen et al. discussed the RF link budget for OAM based single mode communication in the case of a radio wave generated by circular array of isotropic sources \cite{OAMLinkBudget}. The authors highlighted intrinsic limitation for OAM radio communication and equally raised questions on the concepts of gain and signal propagation losses.
Authors of \cite{Cagliero} emphasized that LoS radio OAM transmission through uniform circular arrays (UCAs), similar to what have been used in \cite{OAMPatch1,OAMPatch2,SpinelloARXIV,RadialUCA}, can only deliver spectral efficiency similar to the traditional spatial multiplexing approach only if a perfect alignment is established. Any small misalignment in the positions of the arrays will lead to a significant performance degradation.  
Furthermore, through the aperture antenna theory, Morabito et al. found that OAM antennas are not convenient for long distance transmission \cite{OAMAntennas}. The problem lays with the far-field zone where OAM channel multiplication is unfeasible for beams with a large $\lvert\ell\lvert$ index. Nevertheless, if the receiving antenna is not large enough to detect a maximum of power, the use of OAM modes has a huge impact on the power of the system.  
\section{OAM Perspectives}
Here we review the main perspectives of OAM in various application areas.
\subsection{Towards a Complete Mode Set}
\begin{figure}[!t]
\centering
\includegraphics[width=3in]{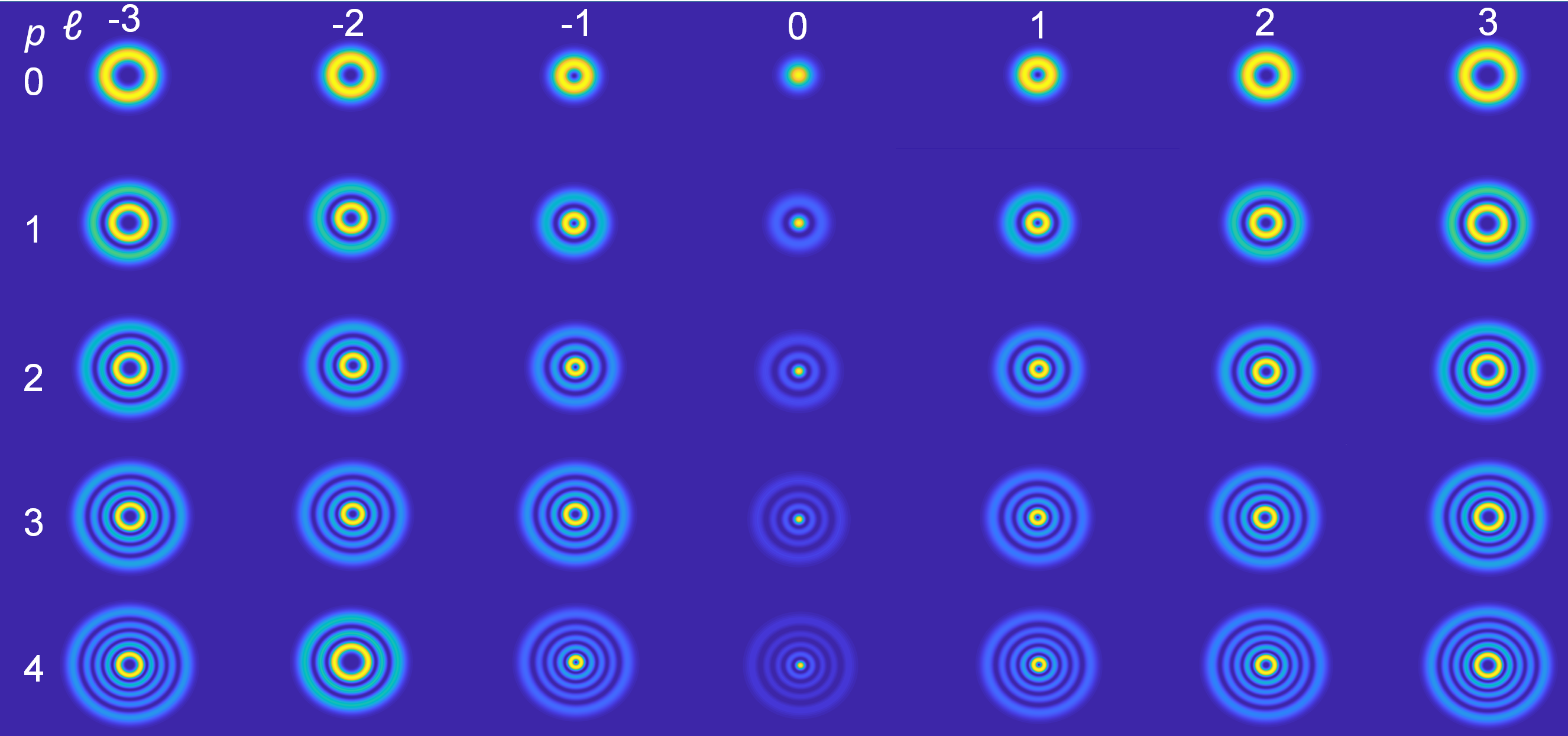}
\caption{Intensity profiles of Laguerre Gaussian modes.}
\label{LGprofiles}
\end{figure}

Very recently, it was pointed out that OAM multiplexing is not an optimal technique for free-space information encoding and that OAM itself does not increase the bandwidth of optical communication systems \cite{ChenSREP16,ZhaoNATPH15}. 
Miller in \cite{PNASMiller} demonstrated through calculations, based on the singular value decomposition (SVD) of the coupling operator between transmitting and receiving sources \cite{SVDMiller}, that there are better optimal choices than using different OAMs to increase the capacity of optical communication. The main output of the study is that OAMs could be outperformed by any complete modal basis to scale the capacity of a communication system. 
In fact, OAM is only a subspace of the full Laguerre Gaussian mode basis. The LG beams form a complete orthonormal basis and are characterized by the azimuthal index $\ell$ and radial index $p$, where the former is responsible for the OAM. The LG modes are described by \cite{Siegman}:
\begin{equation}
\label{LGfield}
\begin{split}
E_{(p,\ell)}^{LG}(r,\phi,z)&=\frac{1}{w(z)}\sqrt{\frac{2p!}{\pi(|\ell|+p)!}}\exp(i(2p+|\ell|+1)\Phi(z))\\
&\times\left (\frac{r\sqrt{2}}{w(z)}\right)^{|\ell|}L_{p}^{|\ell|}\left (\frac{2r^{2}}{w^{2}(z)}\right)\\
&\times\exp\left(-\frac{ikr^{2}}{2R(z)}-\frac{-r^{2}}{w^{2}(z)}+i\ell\phi\right),
\end{split}
\end{equation}
where $\Phi(z)$ represents the Gouy phase, $w(z)$ is the beam spot size parameter, $R(z)$ is the radius of the beam curvature and $L_{p}^{|\ell|}$ are the generalized Laguerre polynomials. A set of the theoretical intensity profiles of LG modes with different radial and azimuthal components is depited in Fig.~\ref{LGprofiles}.\newline
Different encoding schemes have been implemented to demonstrate how to transmit information on LG modes incorporating two degrees of freedom, reaching a total number of 105 modes over 3 wavelengths in \cite{TrichiliSREP16}. Researchers have been able to achieve this, through a holographic approach, using over 100 modes encoded on a single computer generated hologram, across a wide range of wavelength, in an independent manner. The proof of principle experiment was mainly based on SLMs and coupled charge device (CCD) cameras.\newline
According to \cite{TrichiliSREP16}, in free space, LG mode multiplexing has a comparable complexity to OAM multiplexing. This is due to the fact that the propagation dynamics, including beam size, phase shift and divergence are ruled by an indicator known as the beam quality factor $M^{2}=2p+\lvert\ell\lvert+1$. Therefore, $M^2$ can be considered as mode index, and a LG mode with non-zero radial index can propagate in identical manner with an OAM mode having the same beam quality factor. For example LG$_{05}$, LG$_{21}$ and LG$_{13}$ will undergo the same propagation since each mode has the same $M^2$ value of 6.\newline
\indent Additional experiments reported the information encoding using OAM modes with non-null radial components \cite{Xie,LossLGMitigation}. In \cite{Xie}, a 200 Gbit/s transmission was reported using two distinct LG modes with distinct radial components achieving a BER below the forward error correction (FEC) limit. In \cite{LossLGMitigation}, it has been shown that LG beams can significantly contribute to solving the problem of the limited receiving aperture sizes which in conjunction with OAM divergence can lead to a significant loss of power. The LG modes with non-null radial indices can be detected using the same detection techniques for OAM with no radial components modes  through two-dimensional computer generated holograms \cite{TrichiliSREP16}. Novel LG mode de-multiplexers have been also proposed \cite{LGsorterFontaine,BoydSorter}. Fontaine and co-workers demonstrated a device that is able to simultaneously  detect the $\ell$ and $p$ indices of up to 210 LG modes \cite{LGsorterFontaine}. Authors of \cite{BoydSorter} proposed a nearly crosstalk-free scalable LG mode sorter based on a previously demonstrated radial mode sorter \cite{MirSorter} cascaded with an $\ell$-dependent phase shifter and OAM index sorter.

\subsection{Atmospheric Turbulence Effects Mitigation}
\label{Turbulence}
Several techniques have been proposed to mitigate turbulence distortions in free space. Turbulence mitigation can be either performed at the beam level through adaptive optics (AO) and wavefront sensing techniques, or at the signal processing level through channel coding and channel equalization techniques.
Theoretical studies performed in the 70s showed that atmospheric turbulence has a reciprocal nature \cite{Reciprocity1,Reciprocity2}, meaning that two laser beams propagating in opposite directions experience the same atmospheric turbulence distortions. Atmospheric turbulence reciprocity was later confirmed experimentally \cite{Reciprocity3,Reciprocity4}. Through reciprocity, a pilot signal from a receiver can be used at a transmitter to provide channel state information (CSI), and based on this, various strategies may be adopted at the transmission terminal, including adaptive modulation, varying bit rate, and beam phase and amplitude pre-compensation \cite{Reciprocity1}. Reciprocity is also a key characteristic for adaptive optics pre-compensation. By exploiting reciprocity, systems can cope with the complexity and latency issues that limit FSO communications \cite{Reciprocity5}.\newline
\subsubsection{Adaptive Optics}
AO are a technology used to improve the performance of optical systems by reducing the effect of wavefront distortions. It aims at correcting the deformations of an incoming wavefront by deforming a mirror, or by controlling a liquid crystal array in order to compensate for aberrations. AO have been used for astronomy \cite{AOpticsX1}, imaging \cite{AOpticsX2} and microscopy \cite{AOpticsX3}. In optical communication, AO have been found to be applicable in free space by correcting the wavefront of the beam, over which information is encoded, from atmospheric turbulence before being directed to the photodetector \cite{AOptics1,AOptics2,AOptics3}. \newline
\indent The use of adaptive optics is a potential solution to truly implement OAM-based communications in order to undo the propagation effects, mainly over free space links \cite{ShackHartmann,RenOL14}. Authors of \cite{ShackHartmann} used two AO correction approaches to compensate for atmospheric distortions on OAMs with the first based on the Shack-Hartmann algorithm and the second based on a phase correction method for the OAM beams. In \cite{RenOL14}, authors used a Gaussian beam on one polarization as a probe beam to estimate wavefront distortions caused by laboratory emulated atmospheric turbulence and derive correction patterns to efficiently compensate multiple OAM beams on the orthogonal polarization.\newline 
\indent Reciprocity in an OAM communication system has been significantly exploited in \cite{RenOPTICA14} where post- and pre-compensation of OAM beams have been reported. A stochastic-parallel-gradient-descent algorithm, based on Zernike polynomials, has been further used to correct the phase of distorted OAM modes \cite{CorrectionAlgorithm}.\newline
AO have also been found useful in the case of multicasting transmissions, as shown in\cite{MulticastingAO} where an adaptive feedback correction method was adopted to compensate for 4-fold and 8-fold distorted OAM multicasting channels. In this study, the authors successfully suppressed the inter-state crosstalk to reduce power fluctuations between the multicasted OAM channels and they reported a BER performance improvement. \newline
\subsubsection{Coding and Equalization Techniques}
An alternative way to adaptive optics is to perform the correction at the digital domain, using channel coding and equalization techniques. 
An experimental low-density parity check (LDPC)-coded OAM scheme was proposed to compensate for free space propagation effects as reported in\cite{CodingLDPC1,CodingLDPC2}.\newline
A MIMO digital processing equalization scheme associated with heterodyne detection was shown to mitigate turbulence-induced crosstalk for 4 OAM beams carrying 20 Gbit/s QPSK signals \cite{CodingFSOMIMO}. In \cite{SunOPEX17}, Sun et al. experimentally investigated the performance of crosstalk mitigation for 16-QAM OFDM signals over OAM multiplexed FSO communication links, using the pilot-assisted least square (LS) algorithm. It was found to improve the error vector magnitude (EVM) and the BER of the system. In \cite{OAMSpaceCoding}, Zhang el al. investigated the impact of space time coding on the performance of OAM MIMO multiplexing scheme in a turbulent free space link and reported that vertical Bell Laboratories layered space-time (V-Blast) and space-time block codes (STBC) can significantly help undo the effect of atmospheric distortions and noise. Amhoud et al. suggested the use of the Golden and the Silver space time codes to mitigate the effect of atmospheric turbulence \cite{OAMSelection}.\newline
\indent To mitigate the atmospheric turbulence for OAM links, it is also possible to combine channel coding and adaptive optics wavefront correction, as shown in \cite{CodingAO,500GBLDPC,Djordjevic2}. The authors of \cite{CodingAO} first utilized a Shack-Hartmann OAM wavefront correction technique to reduce the turbulence induced intermodal crosstalk and then corrected the residual single OAM random errors using the BCH error correction codes. Qu and Djordjevic also combined AO based wavefront correction with LDPC coding to compensate for intermodal crosstalk induced by atmospheric turbulence \cite{500GBLDPC,Djordjevic2}.\newline
\subsubsection{Modal Diversity}
An additional approach to overcome the effect of atmospheric turbulence, known as modal diversity, has been recently proposed \cite{SafariDiversity,ModalDiversity}. The concept is different from the conventional diversity schemes where signals from multiple spatially-separated transmitters are sent through different paths and collected by multiple receivers so each beam experiences different atmospheric turbulence conditions \cite{TraditionalDiversity}. Modal diversity consists of encoding the same information on different modes co-propagating  through the same turbulent FSO channel and without any requirement on the physical spatial separation between the transmitting and receiving apertures. The principle relies on the fact that distinct spatial modes undergo different distortions. In \cite{SafariDiversity}, the authors simulated modal diversity using well-spaced OAM modes ($\ell=1$, $\ell=8$ and $\ell=15$) and reported a significant performance improvement through diversity which was found to reduce the outage probability of the FSO communication link. Cox and co-workers used orthogonal OAM modes with non-null radial indices and Hermite Gaussian modes to transfer information through atmospheric turbulence with high fidelity, using two transmitters and a single receiver \cite{ModalDiversity}. The authors reported a 54\% improvement in the uncoded BER without changing the transmitted power. M. Li proposed in \cite{MingLi2} an adaptive mode adjustment scheme to compensate for the impact of atmospheric turbulence. The core idea behind the proposed scheme is to adjust the spacing between the multiplexed modes at the transmitter side if the intermodal crosstalk is higher than a specific threshold after demultiplexing at the receiver side. \newline

 Yang et al. proposed a novel technique that does not incorporate adaptive optics, channel coding or modal diversity, to mitigate turbulence \cite{YangOE17}. The idea is to use a Gaussian beam as a local oscillation light, together with an OAM beam acting as a light signal and numerical results showed improvement in the BER system performance. However, such a technique could possibly not be cost-effective especially for free space applications  due to the expensive cost of coherent detecors \cite{HuangIEEETWC18}.\newline
More details on atmospheric turbulence compensation techniques can be found in \cite{ATOAM}. 
\subsection{Machine Learning-Based OAM Recovery Approach}
\label{MachineLearning}
In section \ref{Turbulence}, we reviewed the AO and DSP-based atmospheric turbulence mitigation approaches with the goal of improving the performance of FSO communication systems. One approach is to use a machine learning algorithm for the OAM intensity pattern recognition, before retrieving the initially transmitted information, and without compensating for channel turbulence through AO or signal processing. Krenn et al. used an artificial neural network algorithm to distinguish the intensity patterns of 16 petal shaped OAM modes $(\ell=0,\ell=\pm1,\cdots,\ell=\pm15)$ \cite{KrennNJP14}. A similar algorithm was used to identify OAM beam profiles, with an accuracy exceeding 80\%, over a free space propagation distance of 143 km \cite{143km}. Another study highlighted the potential of using a neural network machine learning algorithm to identify OAM modes with high $\ell$ charges \cite{OAMMachineLearning1}. Doster et al. demonstrated that  a convolutional neural network (CNN) deep learning-based method could detect the combination of OAM states with a precision exceeding 99\% at high atmospheric turbulence levels \cite{OAMDetectionMLearning}. In \cite{OAMMachineLearning1}, Knutson performed a comparative study of several machine learning algorithms to identify OAM mode intensity patterns. Based on deep neural networks, a high-fidelity identification technique for detecting one mode among a set of 100 LG modes is demonstrated in \cite{OAMMachineLearningComp1}. A machine learning based adaptive demodulation techniques for OAM index-based communication was also proposed in\cite{OAMMachineLearning2}.
In \cite {OAMDetectionPRecognition}, Park et al. used a new theory, developed in\cite{Theory} and suggesting a link between the turbulence and the photon transport through the continuity equation, to describe a method that utilizes a ``shallow'' learning method in replacement of the previously algorithms described. The authors also reported the use of only a fraction (1/90) of the computational time, with a similar accuracy as that obtained with the other machine-learning algorithms, making their technique suitable for high bit rate optical communications. In \cite{OAMMachineLearning3}, Li et al. presented a technique for the detection of atmospheric turbulence and adaptive modulation using CNN achieving high accuracy for different numbers of LG modes. Lohani and Glasser designed a feedback scheme based on CNNs to pre-correct OAM profiles at the transmitter end before propagating through turbulent atmosphere, which significantly enhanced the identification process at the receiver end \cite{OAMMachineLearningComp2}.
Authors in \cite{AlignOAMML} proposed the use of CNN to detect OAM beams subject to turbulence and misalignment errors between the transmitter and the receiver.
\newline
\indent The use of machine learning algorithms for identifying the shape of  OAM beams can be further exploited in optical fiber and underwater transmissions. 
\subsection{Beam Tracking}
For the outdoor implementation of FSO, building sway due to wind, thermal expansion and minor earthquakes may cause a slight deviation of the beam \cite{ArnonIEEETWC03}, when propagating between two communicating terminals. The OAM mode detection is very sensitive to alignment and pointing errors. One way to compensate for beam displacements, in order to obtain a reliable mode decoding, is to perform beam tracking. This has attracted considerable attention lately. Beam tracking techniques have been widely used for Gaussian modes and have been found to be convenient for increasing system performance \cite{Tracking1}. Recently, beam tracking has been introduced to OAM wireless optical communication \cite{TrackingOAM1,TrackingOAM2}. OAM beam tracking operations have first been successfully applied to free space communication between two ground terminals relayed by an unmanned aerial vehicle (UAV) carrying a retro-reflector achieving a transmission bit rate of 80 Gbit/s with two OAM channels \cite{TrackingOAM1}. Two tracking sub-systems using components to detect the position and angle of the beam and feedback controllers were harnessed. A Gaussian beacon beam was used to improve the tracking of the OAM beam, which can be affected by a misalignment between the transmitter and the receiver. A beacon-less OAM beam tracking scheme was later reported, showing a similar performance to tracking methods with a Gaussian beacon \cite{TrackingOAM2}.\newline
\indent Tracking the beam has not been only restricted to FSO but was also used in underwater communication.
When establishing the water-air-water communication link, Wang et al. were able to maintain an aligned optical path by tracking the beam displacement caused by a change of the water level \cite{WaterAirWater}. While establishing the NLoS underwater communication link that is based on the total internal reflection enabled by the air-water interface, Zhao et al. also maintained the alignment of the optical path by tracking the beam displacement caused by tide and surface waves, using two feedback loops \cite{NLOSOAM}.

\subsection{Underwater Turbulence Effects Mitigation}
In \cite{RENSREP16}, a thermal gradient induced turbulence was mitigated using a multi-channel equalization method to ensure the recovery of information carried by two OAM signals.\newline
AO have been equally used in an underwater environment to compensate for the turbulence caused by thermal gradient \cite{UnderwaterAO}. However, such a technique has not yet been tested on OAM-based UWOC. \newline
\indent To overcome the effect of bubbles, it is possible to use Bessel Gaussian modes, instead of OAM LG modes, since they can self-heal while being obstructed during propagation. This may enhance the performance of the UWOC, as shown in \cite{ZhaoOL17}.
The underwater turbulence mitigation techniques are still limited in comparison with free space; this may be partially explained by the novelty of the UWOC.
\subsection{OAM Maintaining Fibers}
LP modes intrinsically couple to each other, during propagation, due to the superposition of non-degenerate (vector) fiber modes, rather than exact mode solutions. In section \ref{FiberPotential}, we show that OAM modes can efficiently propagate in conventional FMF and MMF, and in specially designed fibers. Different fiber technologies have been used in order to efficiently support OAMs including photonic crystal fibers (PCFs) \cite{PCFOAM2}, inverse-parabolic graded-index fibers \cite{ParabolicOAM}, air-core fibers \cite{AirCoreOAM1,AirCoreOAM2}, multi-ring fibers \cite{RingFiberx1,RingFiberx2,MCFOAMRing,MCFOAM1}, and multicore supermode fibers \cite{MCFOAM2}.
The key to allow the propagation of OAM modes in optical fibers is to ensure a separation between the effective refractive indices of different mode groups of more than $10^{-4}$ \cite{OAMFiberAnaly}. Such a small difference is hard to obtain but could break the weakly guiding approximation allowing the propagation of OAM modes.
 In Table \ref{OAMFibers}, we summarize the main key characteristics of the designed and manufactured OAM maintaining fibers in the literature. \newline
Light propagation through a material is minimized in air-core fibers, which significantly reduces the intra- and inter-modal nonlinear effects.
Note that several other fibers that support twisted light were successfully introduced, but for applications other than communication \cite{PCFOAM1}.
\begin{table*}[!t]
\renewcommand{\arraystretch}{1.3}
\caption{Summary on the designed OAM maintaining fibers}
\label{OAMFibers}
\centering
\begin{tabular}{|>{\columncolor[gray]{0.8}}l|l|p{0.4\textwidth}|}
\hline
\rowcolor[gray]{0.8}
Reference & Fiber Type &Key Characteristics\\
\hline
\cite{PCFOAM2} &Photonic Crystal&\vspace{-\topsep}\begin{itemize}[leftmargin=*]   
\item Designed fiber supporting 26 OAM modes
\item Low confinement loss and small nonlinear coefficient
\item Flat chromatic dispersion
\item Suitable for future mode division multiplexing systems
  \end{itemize}\vspace{-\topsep}\\
\hline
\cite{ParabolicOAM} & Inverse Parabolic Graded-Index&\vspace{-\topsep}\begin{itemize}[leftmargin=*]   
\item Manufactured fiber, supporting 2 modes for 1.1 km

\item Large effective refractive index difference between modes
\item Suitable for practical short distances MDM transmission
  \end{itemize}\vspace{-\topsep}\\
\hline
\cite{AirCoreOAM1}&Air-Core&\vspace{-\topsep}\begin{itemize}[leftmargin=*]   
\item Manufactured a hollow core fiber, supporting 36 OAM states
\item Very high loss up to few dBs per meters 
\item Cannot be used for communication
  \end{itemize}\vspace{-\topsep}\\
\hline
\cite{AirCoreOAM2}&Air-Core&\vspace{-\topsep}\begin{itemize}[leftmargin=*]   
\item Manufactured fiber, supporting 12 OAM modes
\item High purity for 8 OAMs after 1 km
\item Promising for communication over short distances 
  \end{itemize}\vspace{-\topsep}\\
\hline
\cite{MCFOAMRing}&Multicore/Multi-Ring&\vspace{-\topsep}\begin{itemize}[leftmargin=*]   
\item Designed fiber, having 7 cores with each supporting 18 OAM modes
\item Inter-ring crosstalk lower than -30 dB for a 100 km fiber
\item Low inter-modal crosstalk for the wavelength range of 1520-1580 nm
  \end{itemize}\vspace{-\topsep}\\
\hline
\cite{MCFOAM1}&Multicore/Multi-Ring&\vspace{-\topsep}\begin{itemize}[leftmargin=*]   
\item Designed fiber, having 19 cores with each supporting 22 OAM modes
\item Low ring ellipticity and bending induced crosstalk levels
\item Might enable Pbit/s transmission capacities
  \end{itemize}\vspace{-\topsep}\\
\hline
\cite{RingFiberx1,RingFiberx2}&Multi-Ring&\vspace{-\topsep}\begin{itemize}[leftmargin=*]   
\item Manufactured fiber, supporting multiple OAM
\item Low inter- and intra-modal OAM group crosstalk
\item Small GVD slope around the C band (1535 nm-1565 nm)
\item Tested in different configurations 
\end{itemize}\vspace{-\topsep}\\
\hline
\cite{RingCore}&Ring Core Graded Index&\vspace{-\topsep}\begin{itemize}[leftmargin=*]   
\item Manufactured fiber, supporting 5 mode groups
\item Large effective refractive index separation between mode groups
\item Low inter-modal crosstalk
\item Supported an 8.4 Tbit/s transmission over 18 km
\item Smooth demultiplexing without the need to MIMO DSP
\end{itemize}\vspace{-\topsep}\\
\hline
\cite{MCFOAM2}&Multicore Supermode&\vspace{-\topsep}\begin{itemize}[leftmargin=*]   
\item Manufactured 6 core fiber, supporting 12 eigenstates
\item Low mode dependent loss and nonlinear coefficient
  \end{itemize}\vspace{-\topsep}\\
\hline
\end{tabular}
\end{table*} 
A further theoretical study reported in \cite{GradedIndexOAMFiber} proposed a graded index fiber for the propagation of 99.9\% purity OAM. The fiber supports 10 OAM modes with low inter-modal crosstalk of -30 dB and low dispersion with a GVD coefficient which equals to -35 ps/(km.nm).  
Through the use of a recirculating fiber loop, Gregg et al. in \cite{Gregg} demonstrated low propagation losses of multiple OAM modes over 13.4 km, in an air-core fiber with the same characteristics as those highlighted in \cite{AirCoreOAM2}. The authors reported a considerable decrease of the inter-modal crosstalk due to an increase of the refractive index separation.
Such optical fibers could be very promising for data center-guided interconnects \cite{FiberDataCenter}. The ring-core graded index fiber, reported in \cite{RingCore}, and over which an 18 km OAM-WDM transmission with a low crosstalk was demonstrated, could be also used for interconnect applications. \newline
\indent Encoding data over long-distance OAM-maintaining fibers requires developing new OAM signal amplification techniques. Some recent studies proposed inline OAM amplifiers \cite{OAMAmplification1,OAMAmplification2}. In \cite{OAMAmplification1}, Kang et al. theoretically designed a 12 OAM mode erbium-doped fiber amplifier (EDFA). The proposed fiber has an air-core structure with a refractive index profile matching the one of the used fiber in \cite{AirCoreOAM2}. The gain is quasi-uniform between the OAM modes with the highest differential modal gain (DMG) of 1 dB at 1530 nm. In \cite{OAMAmplification2}, Jung et al. reported the first controllable amplification of low-order OAM modes with $\lvert\ell\lvert=1$ through an air-hole erbium-doped fiber. 
\subsection{Optimized RF Transceiver/Receiver Design}
We start by reviewing the different proposed solutions for OAM generation and detection to overcome the challenges of radio OAM communication. One solution is the design of antenna allowing the generation of high purity focused radio OAM \cite{HuiSREP15}. Another possible way to overcome the divergence in OAM radio communication is to generate OAMs with Bessel structure rather than beams with Laguerre Gaussian profiles \cite{Convergence}. A two-layered Rotman lens was manufactured and utilized to generate high purity five OAM radio states \cite{RotmanLens}. The accurate detection of OAM beams through the observation of the Doppler shift effect is addressed in \cite{VirtualOAM1,VirtualOAM2}. Another technique for the detection of multiple OAM RF states using a single metasurface is modeled in \cite{ChenIEEEAWPL}. An achromatic lens was used to undo the effect of chromatic aberration \cite{AchromaticLens}. The problem of multipath was addressed in \cite{OAMFFT} through the use of a two dimensional fast Fourier transform (FFT). Authors of  \cite{UCAOAM} proposed a method to predict the link budget of OAM links based on UCAs. In Table \ref{OAMRFsolutions}, we summarize the different proposed solutions to OAM radio link challenges described in the literature.\newline
\indent Some of the proposed techniques, presented in Table \ref{OAMRFsolutions}, are focused on suppressing or minimizing the effects, whereas some others aim to overcome the challenges without performing any correction at the beam level.\newline
\begin{table*}[!t]
\renewcommand{\arraystretch}{1.3}
\caption{Summary of the RF OAM proposed solutions}
\label{OAMRFsolutions}
\centering
\begin{tabular}{|>{\columncolor[gray]{0.8}}l|l|p{0.5\textwidth}|}
\hline
\rowcolor[gray]{0.8}
Reference & Adressed Problem(s) &Key Features and Solutions\\
\hline
\cite{HuiSREP15}&Beam divergence&\vspace{-\topsep}\begin{itemize}[leftmargin=*]  
\item  Generated a 60 GHz radio communication link that simultaneously supports two separated channels 
\item A parabolic reflector is used to minimize divergence and to focus the coaxially propagating OAM radio beams
  \end{itemize}\vspace{-\topsep}\\
\hline
\cite{Convergence}&Beam divergence&\vspace{-\topsep}\begin{itemize}[leftmargin=*]   
\item Generated a converging and non-diffracting Bessel-like OAM beams based on phase-engineered metalenses for microwave applications
\item Introduced the possibility to control the OAM beam radius differently from all the previously demonstrated OAM generation techniques
  \end{itemize}\vspace{-\topsep}\\
\hline
\cite{RotmanLens}&Beam purity, divergence&\vspace{-\topsep}\begin{itemize}[leftmargin=*]   
\item Proposed an effective technique to generate multiple OAM beams via an antenna array fed by a two-layer Rotman lens
\item Low cost and easy to implement and integrate with available technologies
  \end{itemize}\vspace{-\topsep}\\
\hline
\cite{VirtualOAM1}&OAM detection&\vspace{-\topsep}\begin{itemize}[leftmargin=*]   
\item Designed a reception technique for OAM millimeter waves consisting of receiving a part of the signal in the time domain rather than receiving the complete signal in the space domain
\item Rotated the OAM waves without altering the polarization of the beams at the transmitter and used a single fixed antenna at the reception 
  \end{itemize}\vspace{-\topsep}\\
\hline
\cite{VirtualOAM2}&OAM detection&\vspace{-\topsep}\begin{itemize}[leftmargin=*]   
\item Used the Doppler shift induced when rotating OAM modes to detect the waves at the reception
\item Digitally rotated the receiving antenna using DSP methods instead of mechanical rotation
  \end{itemize}\vspace{-\topsep}\\
\hline
\cite{ChenIEEEAWPL}&OAM detection&\vspace{-\topsep}\begin{itemize}[leftmargin=*]    
\item Proposed a single metasurface for the detection of multiples OAM modes
\item Modified the transmittance function of the metasurface to maximize the directivity
  \end{itemize}\vspace{-\topsep}\\
\hline
\cite{AchromaticLens} &Bandwidth, chromatic aberration&\vspace{-\topsep}\begin{itemize}[leftmargin=*]   
\item Efficiently converted circularly polarized waves to OAM carrying waves over a broad bandwidth using a multi-layer achromatic metasurface
  \end{itemize}\vspace{-\topsep}\\
\hline
\cite{OAMFFT} &Multipath&\vspace{-\topsep}\begin{itemize}[leftmargin=*]   
\item Proposed a 2D-FFT based transceiver architecture
\item Low complexity algorithm for OAM-OFDM implementation
  \end{itemize}\vspace{-\topsep}\\
\hline
\cite{UCAOAM} &Gain and phase mismatches&\vspace{-\topsep}\begin{itemize}[leftmargin=*]   
\item Proposed a calibration process based on mathematical derivations to estimate, and compensate the gain and phase mismatches between the UCA-based transmitter and receiver simultaneously. 
  \end{itemize}\vspace{-\topsep}\\
\hline
\end{tabular}
\end{table*} 
In order to efficiently exploit the spatial structure of the radio waves, researchers proposed to use distinct OAM states to encode and decode information \cite{OAMRadioDE1,OAMRadioDE2}. In \cite{OAMRadioDE2}, Liu et al. presented a new design of single port antenna that can be used to transmit $\ell=1$ and $\ell=-1$ OAM modes. Data bits 1 and 0 were mapped to $\ell=1$ and $\ell=-1$ OAM modes, respectively. Very recently, a new study proposed enabling the index modulation of the OAM states to convey information, rather than using OAM modes as information carriers to get better performances at the same detection complexity \cite{BasarIEEETWC18}.\newline
\indent The claims about the high complexity of the use of multiple modes in RF communication systems were revisited in \cite{ExperimentalStudy}, in which Zhang et al. demonstrated, through mathematical analysis and experimental validation, that multiple OAM modes could significantly reduce the receiver's complexity, compared to a MIMO single mode communication, without affecting the system capacity.
The limit of the OAM RF link budget estimate reported in \cite{OAMLinkBudget}, was discussed by Cagliero et al.; they provided a new approach to estimate the link budget \cite{OAMBudget2}. The authors took into account the spatial structure of the beam which yielded a correct description of the OAM communication link considering uniform circular antenna arrays. 
Authors \cite{Friis} derived a generalized Friis transmission equation that could be used to estimate the link budget of radio OAMs incorporating UCAs. The equation can be also used to approximate the Rayleigh length of OAM of charge $\ell$. \newline
\indent Zhang et al. claimed that ``the negative arguments'' on OAM radio are only based on comparison of OAM systems that use UCAs to generate twisted waves with MIMO systems \cite{OAMMiMo}. According to the authors, the fact that UCAs are formed with multiple connected antennas fed by beamforming networks leads to the comparison of OAM radio systems with MIMO and therefore suggested the use of OAM modal diversity instead of the orthogonality to increase the transmission capacity of OAM systems.
Multiplying and canceling of OAM modes, using a Cassegrain reflector antenna, was demonstrated in the near-field zone highlighting the potential for using OAM modes in near-field links \cite{ReflectarrayOAM,NearField}. Owing to their special signature in the far-field and with no information transmission limitations, OAM radio beams were suggested for satellite navigation applications \cite{ReflectarrayOAM}.\newline
\begin{table*}[!t]
\renewcommand{\arraystretch}{1.3}
\caption{Summary of impact of OAM multiplexing on the communication key metrics}
\label{KeyMetrics}
\centering
\begin{tabular}{|>{\columncolor[gray]{0.8}}l|p{0.5\textwidth}|}
\hline
\rowcolor[gray]{0.8}
Application area &Key Metrics\\
\hline
Free space optics&\vspace{-\topsep}\begin{itemize}[leftmargin=*]   
\item Scalable transmission capacity and spectral efficiency when independent data streams are encoded on different modes
\item Lower BER and outage probability through the application of modal diversity
\item Higher physical layer security under atmospheric turbulence effect
\item Energy efficiency when applied in data centers
\end{itemize}\vspace{-\topsep}\\
\hline
Optical fiber communication&\vspace{-\topsep}\begin{itemize}[leftmargin=*]   
\item Scalable transmission capacity
\item Lower complexity when MIMO is not required for mode separation at the reception
\item Convergence with optical wireless systems that also use OAMs to convey information without the need for intermediate electrical devices to connect a fiber to an optical wireless channel 
  \end{itemize}\vspace{-\topsep}\\
\hline
Radio Frequency&\vspace{-\topsep}\begin{itemize}[leftmargin=*]   
\item Higher spectral efficiency and transmission capacity in an ideal medium with perfect alignment
\item Special beam signature for radar and imaging application
  \end{itemize}\vspace{-\topsep}\\
\hline
Chip-to-chip&\vspace{-\topsep}\begin{itemize}[leftmargin=*] 
\item Higher transmission capacity
\item Faster processing than electrical circuits
\item All optical functions
  \end{itemize}\vspace{-\topsep}\\
\hline
Acoustic communication&\vspace{-\topsep}\begin{itemize}[leftmargin=*]   
\item Higher transmission capacity and spectral efficiency
\end{itemize}\vspace{-\topsep}\\
\hline
\end{tabular}
\end{table*} 
\section{Open problems and future research directions}
In this section, we summarize the open problems in each of the practical application areas of OAM multiplexing, which could be considered as a future research direction: 
\subsubsection{Free space optics}
Maintaining the alignment between the laser and the external mode generation tool (a spatial light modulator for example) is crucial in FSO OAM links but can be sometimes complicated and not suitable for out-of-laboratory implementation. This can be solved through the use of compact devices to generate OAMs. Another major requirement for an FSO communication is to have an LoS configuration. Connecting shadowed terminals using OAM-FSO requires the use of relay nodes suitable for OAM modes. Excepting the use of mirrors to connect NLoS terminals \cite{TrackingOAM1,WaterAirWater}, using relays has not yet been investigated in communication involving spatial modes. A potential relaying scheme with a \emph{correct and forward} capability could be explored. The idea consists of applying corrections on an incoming beam from a transmitting node using adaptive optics and forward it to a receiving node. Relay nodes can as well be used to reduce the size of the beams that intrinsically tend to diverge while propagating. As we pointed out in section \ref{FSOSubSection}, the theoretical framework of the propagation of OAM  beams through turbulent channel modes and the dependence of the mode indices (of OAM or LG in general) to weather conditions is still missing. Such framework allows the performance evaluation of the system without restoring to expensive and time consuming experimental trials.
\subsubsection{Optical fibers}
We have shown that recently new OAM amplifiers are proposed. However, the proposed demonstrated amplifiers are either not yet being designed experimentally \cite{OAMAmplification1} or still limited to low number of modes \cite{OAMAmplification2}. Therefore, further efforts should be dedicated to develop new amplifiers. Deleterious impact of fiber nonlinearity on OAM beam propagation is little-known and should be explored. Harnessing propagation effects on OAM in fibers could also help to develop new nonlinear devices such as laser sources and optical amplifiers.
\subsubsection{Underwater propagation}
Understanding how the different underwater turbulence scenarios affect the propagation of multiple OAM beams is needed experimentally and theoretically. This will equally pave the road to the development of mitigation strategies either through AO or DSP techniques.
Similarly to free space communication, maintaining the alignment between the generation unit and the laser in an environment is a major concern. This will require devices that could be used inside the water.
\subsubsection{On-chip applications}
Additional small-footprint generation devices are needed to bring OAM to the chip level. Likewise, having a chip-size OAM detector, similar to what have been proposed in \cite{OAMDetectionPCircuit,3DOAMDetector}, is obviously necessary to implement spatial mode-based communication between photonic integrated circuits. 
\subsubsection{RF}
OAM RF communication is still a growing field but full with challenges in particular for far-field applications. Using OAM in near-field applications requires more-in-depth exploration especially in the transmitting and receiving antennas design. 
\section{Discussion}
From what we have presented, we summarized in Table \ref{KeyMetrics} the communication key metrics that could be enhanced through the use of OAMs as information carriers in the different application areas.
From what we have also reviewed, we can see that the biggest advancements in multiple mode transmission are in free space and optical fiber communications. Free space OAM links can provide reliable outdoor connectivity for short and moderate distances. Turbulence and propagation effect mitigation techniques based on AO and DSP will continue to play a fundamental role in reducing the inter-modal crosstalk and improving the system performance. On the other hand, modal diversity could be a cost-effective solution to cope with turbulence effect without the need for complex equalization algorithms and adaptive optics. Similar design considerations to those outlined in \cite{OpticaFSO} are really practical for the deployment of OAM FSO data links helping to choose the mode spacing. Indoor OAM-based links can reduce cable congestion and energy consumption in data centers. The use of OAM modes with non-zero radial component can provide an additional degree of freedom to FSO communication.\newline
\indent Guiding OAM over fibers seems to be very promising, especially after the successes obtained with MMF. This is actually very encouraging to cost-effectively exploit existing MMF installations in most of the major cities. Developing customized integrated OAM laser sources, such as that used in \cite{VCSELOAM,Per1}, suitable for MMF transmission can contribute as well to the deployment of additional OAM-based fiber links in data centers or fiber to the home applications and therefore reducing the cabling and energy consumption. The fact that launching OAM modes in fibers does not require complex MIMO processing is appealing for practical deployments of this technology. With the fast progress in OAM measurement techniques, it is safe to say that commercial OAM demultiplexers that can be used in high-speed communication will be soon available. Additional attention should be given to multimode EDFAs to scale up the transmission distances.\newline 
\indent OAM switching and routing functions will find potential uses in all-optical photonic circuits. Progress on OAM on-chip is slow but sure. OAM-based chips will fasten processing time in computers as well as servers.\newline
\indent For underwater turbulence, we must emphasize that studies investigating the effects of turbulence are still limited in number, and the considered scenarios are far from real-life conditions. Differently from free space and optical fibers, very few studies showed the simultaneous transmission of multiple optical OAM beams underwater which requires more research to quantify the crosstalk between the co-propagating modes. Real life deployment should be also carried out to test the practicability of establishing multiple OAM underwater links.\newline
\indent There is a huge debate on the practicability of OAM radio links and how to face divergence problems without increasing the system power penalty. In addition to experimental and theoretical proofs, real life deployment of OAM radio will not have much interest for long-distance transmissions, due to the fact that the first RF OAM idea promoters \cite{TamburiniNJP12} showed, through calculations published in \cite{SDMOAMRadio}, that spatial mode multiplexing had a limited contribution in comparison with state-of-the-art MIMO systems. However, several researchers stressed the fact that, the particular spatial profile of OAM radio waves could be used to perform radar and satellite localization operations. \newline 
\indent Even after addressing all the technical issues, the cost of deploying OAM communication systems remains a challenge. Cost is a very important metric for setting a communication system and consumer applications may help reducing prices. Cost will play a fundamental role mainly for fiber implementation because commercially available OAM fibers are expensive, and are currently only suitable for laboratory testbed experiments. Having OAM-shaped laser sources will not only reduce the cost of the mode generation techniques, but also the equipment density in the work places.

\section{Conclusion}
The gigantic demand of transmission capacity might push the actual communication systems into a bandwidth bottleneck. Communication using spatial mode of light is viewed as a key solution to cope with the foreseen capacity crunch. Multiplexing spatial modes, in particular OAMs, has been demonstrated over FSO links, optical fibers, and underwater communication links. OAM modes can be further used to perform fast switching and routing functions in optical networks. Bringing OAM to the chip level in photonic circuits is also possible. Additional research has proven the feasibility of communication using multiple OAM radio beams. Besides this, acoustic twisted OAMs are capable of conveying independent information streams to push the acoustic channel capacity.\newline
Several challenges must be overcome in order to implement OAM in communication. Analog and digital approaches are able to efficiently mitigate atmospheric turbulence effects. AO can be used to perform corrections at the beam level, whereas coding, equalization methods as well as machine learning-based algorithms are used to correct optical signals at the signal processing level. The same correction techniques that work well for FSO can be used to compensate for underwater propagation effects. Beam tracking methods are effective against pointing errors in wireless multi-modal communications either in free space or underwater. Novel optical fiber designs are proposed to enhance the OAM stability and maximize transmission data rates with minimal propagation effects. Several major limitations associated with the wave structure affect the performance of OAM in radio communication for long-haul transmission. However, radio OAM for near-field application is promising and requires more research efforts.\newline
\indent The use of the full LG mode basis can deliver higher capacity to OAM-based communication systems. Using mode basis such as the HG and vortex vector basis should be also considered.  
 \section*{Acknowledgment}
Figure [4] was produced by Xavier Pita, scientific illustrator at King Abdullah University of Science and Technology (KAUST). 
 \section*{Acronyms}
\noindent\textbf{AO} Adaptive Optics\newline
\textbf{AUV} Autonomous Underwater Vehicle\newline
\textbf{BER} Bit Error Rate\newline
\textbf{CCD} Charge Coupled Device\newline
\textbf{CD} Chromatic Dispersion\newline
\textbf{CGH} Computer Generated Hologram\newline
\textbf{CMOS} Complementary Metal Oxide Semiconductor\newline
\textbf{CNN} Convolutional Neural Network\newline
\textbf{CSI} Channel State Information\newline
\textbf{DFT} Discrete Fourier Transform\newline
\textbf{DGD} Differential Group Delay\newline
\textbf{DMD} Digital Micro-mirror Device\newline
\textbf{DMG} Differential Modal Gain\newline
\textbf{DMGD} Differential Mode Group Delay\newline
\textbf{DQPSK} Differential Quadrature Phase Shift Keying\newline
\textbf{DSP} Digital Signal Processing\newline
\textbf{EDFA} Erbium Doped Fiber Amplifier\newline
\textbf{EG} Exponential-Gamma\newline
\textbf{EM} Electromagnetic\newline
\textbf{EVM} Error Vector Magnitude\newline
\textbf{FEC} Forward Error Correction\newline
\textbf{FFT} Fast Fourier Transform\newline
\textbf{FH} Frequency Hopping\newline
\textbf{FMF} Few Mode Fiber\newline
\textbf{FSO} Free Space Optics\newline
\textbf{FWM} Four Wave Mixing\newline
\textbf{GGD} Generalized Gamma Distribution\newline
\textbf{GVD} Group Velocity Dispersion\newline
\textbf{HE} Hybrid Electric\newline
\textbf{HG} Hermite Gaussian\newline
\textbf{LCD} Liquid Crystal Display\newline
\textbf{LD} Laser Diode\newline
\textbf{LDPC} Low-Density Parity Check \newline
\textbf{LED} Light Emitting Diode\newline
\textbf{LG} Laguerre Gaussian\newline
\textbf{LoS} Line of Sight\newline
\textbf{LP} Linearly Polarized\newline
\textbf{LS} Least Square\newline
\textbf{MCF} Multicore Fiber\newline
\textbf{MDM} Mode Division Multiplexing\newline
\textbf{MEMS} Microelectromechanical Systems\newline
\textbf{MH} Mode Hopping\newline   
\textbf{MIMO} Multi-Input-Multi-Output\newline
\textbf{MMF} Multimode Fiber\newline
\textbf{NLoS} Non-Line-of-Sight\newline
\textbf{NRZ} Non Return to Zero\newline
\textbf{OAM} Orbital Angular Momentum\newline
\textbf{OFDM} Orthogonal Frequency Division Multiplexing\newline
\textbf{OOK} On-Off-Keying\newline
\textbf{PAM} Pulse-Amplitude Modulation\newline
\textbf{PCF} Photonic Crystal Fiber\newline
\textbf{PLS} Physical Layer Security\newline
\textbf{PDM} Polarization Division Multiplexing\newline
\textbf{PMD} Polarization Mode Dispersion\newline
\textbf{PoC} Proof-of-Concept\newline
\textbf{PPLN} Periodically Poled Lithium Niobate\newline
\textbf{QAM} Quadrature Amplitude Modulation\newline
\textbf{QPSK} Quadrature Phase Shift Keying\newline
\textbf{RF} Radio Frequency\newline
\textbf{ROV} Remotely-operated Vehicle\newline
\textbf{SAM} Spin Angular Momentum\newline
\textbf{SAR} Synthetic Aperture Radar\newline
\textbf{SBS} Stimulated Brillouin Scattering\newline
\textbf{SDM} Space Division Multiplexing\newline
\textbf{SER} Symbol Error Rate\newline
\textbf{SLM} Spatial Light Modulator\newline
\textbf{SM} Spatial Modulation\newline
\textbf{SMF} Single Mode Fiber\newline
\textbf{SNR} Signal-to-Noise Ratio\newline
\textbf{SPM} Self Phase Modulation\newline
\textbf{SPP} Spiral Phase Plate\newline
\textbf{SR} Strehl Ratio\newline
\textbf{SRS} Stimulated Raman Scattering\newline
\textbf{STBC} Space-Time Block Codes\newline
\textbf{SVD} Singular Value Decomposition \newline
\textbf{TE} Transverse Electric\newline
\textbf{TM} Transverse Magnetic\newline
\textbf{TOD} Third Order Dispersion\newline 
\textbf{UAV} Unmanned Aerial Vehicle\newline
\textbf{UCA} Uniform Circular Array\newline
\textbf{UWOC} Underwater Wireless Optical Communication\newline
\textbf{V-BLAST} Vertical-Bell Laboratories Layered Space-Time\newline 
\textbf{VCSEL} Vertical-Cavity Surface-Emitting Laser\newline
\textbf{WDM} Wavelength Division Multiplexing\newline
\textbf{XPM} Cross-phase Modulation\newline

%
%
%

%
%
%
%
\begin{IEEEbiography}[{\includegraphics[width=1in,height=1.25in]{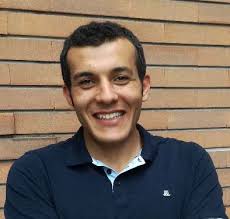}}]{Abderrahmen Trichili}
received his engineering degree and PhD degree in Information and Communication Technology from the Engineering School of Communication of Tunis (Tunisia) in 2013 and 2017, respectively. He is currently a Postdoctoral Fellow at Computer, Electrical and Mathematical Sciences \& Engineering at KAUST University. His current areas of interst include space division multiplexing, orbital angular momentum multiplexing, free space optical communication, and underwater wireless optical communication.  
\end{IEEEbiography}
\begin{IEEEbiography}[{\includegraphics[width=1in,height=1.25in]{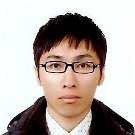}}]{Ki-Hong Park}
(S’06–M’11) received the B.Sc. degree in electrical, electronic, and radio engineering, and the M.S. and Ph.D. degrees from the School of Electrical Engineering, Korea University, Seoul, South Korea, in 2005 and 2011, respectively. Since 2011, he has been a Post-Doctoral Fellow in electrical engineering with the Division of Computer, Electrical, and Mathematical Science and Engineering, King Abdullah University of Science and Technology, Thuwal, Saudi Arabia. He received a SABIC Postdoctoral Fellowship in 2012. His research interests are broad in communication theory and its application to the design and performance evaluation of wireless communication systems and networks. On-going research includes the application to MIMO diversity/beamforming systems, cooperative relaying systems, and physical layer secrecy.
\end{IEEEbiography}
\begin{IEEEbiography}[{\includegraphics[width=1in,height=1.25in]{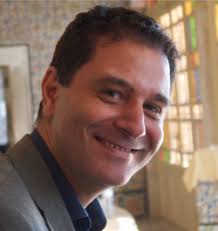}}]{Mourad Zghal}
is a Professor at the University of Carthage, Engineering School of Communication of Tunis (Sup'Com). He received his Ph.D in electrical engineering from the Tunis El-Manar University, National Engineering School of Tunis in 2000. He has published more than 150 papers in areas that include integrated nonlinear optical devices, design and characterization of photonic crystal fibers, nonlinear propagation of ultra-short laser pulses, and mode division multiplexing. He has served at numerous program or steering committee of international scientific conferences. Dr. Zghal has been awarded the 2008 ICO/ICTP Gallieno Denardo prize and is Fellow of OSA and Fellow of SPIE.
\end{IEEEbiography}
\begin{IEEEbiography}[{\includegraphics[width=1in,height=1.25in]{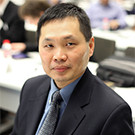}}]{Boon S. Ooi}
 is a Professor of Electrical Engineering at KAUST. He is also the Director of KACST - Technology Innovation Center (TIC) for Solid-State Lighting. Professor Ooi received the B.Eng. and Ph.D. degrees in electronics and electrical engineering from the University of Glasgow (Scotland, U.K) in 1992 and 1994, respectively. He joined KAUST from Lehigh University (Pennsylvania, USA) where he held an Associate Professor position and headed the Photonics and Semiconductor Nanostructure Laboratory. In the U.S., his research was primarily funded by the National Science Foundation (NSF) and Department of Defense and the Army Research Office. In KSA, major funding support for his research is from King Abdulaziz City for Science \& Technology (KACST), Saudi Aramco, SABIC and Qatar National Research Fund (QNRF).
\end{IEEEbiography}
\begin{IEEEbiography}[{\includegraphics[width=1in,height=1.25in]{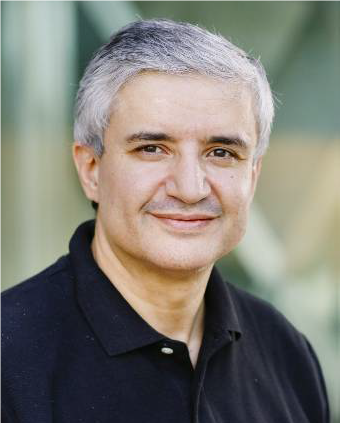}}]{Mohamed-Slim Alouini}
(S'94-M'98-SM'03-F'09) was born in Tunis, Tunisia. He received the Ph.D. degree in Electrical Engineering from the California Institute of Technology (Caltech), Pasadena, CA, USA, in 1998. He served as a faculty member in the University of Minnesota, Minneapolis, MN, USA, then in the Texas A\&M University at Qatar, Education City, Doha, Qatar before joining King Abdullah University of Science and Technology (KAUST), Thuwal, Makkah Province, Saudi Arabia as a Professor of Electrical Engineering in 2009. His current research interests include the modeling, design, and performance analysis of wireless communication systems.
\end{IEEEbiography}

\end{document}